\documentclass[11pt,a4paper]{article}
\usepackage{silence} 
\usepackage{jheppub}
\usepackage{graphicx}
\usepackage{xcolor}
\usepackage{orcidlink} 
\usepackage[makeroom]{cancel}
\usepackage{enumitem}
\usepackage{amssymb,amsmath,mathrsfs,bm,bbm,mathtools} 
\usepackage{amsthm}
\definecolor{ao(english)}{rgb}{0.0, 0.5, 0.0}
\hypersetup{colorlinks=true,linkcolor=blue,urlcolor=blue,citecolor=ao(english)}

\newcommand\nn{\nonumber\\}

\newcommand{\bma}{\left(\begin{array}}
\newcommand{\ema}{\end{array}\right)}
\newcommand{\be}{\begin{equation}}
\newcommand{\ee}{\end{equation}}
\newcommand{\ben}{\begin{equation*}}
\newcommand{\een}{\end{equation*}}
\newcommand{\ba}{\begin{eqnarray}}
\newcommand{\ea}{\end{eqnarray}}
\newcommand{\ban}{\begin{eqnarray*}}
\newcommand{\ean}{\end{eqnarray*}}
\newcommand{\bs}{\begin{subequations}}
\newcommand{\es}{\end{subequations}}
\newcommand{\bc}{\begin{center}}
\newcommand{\ec}{\end{center}}

\newcommand{\Pl}{{\text{\tiny Pl}}}
\newcommand{\lp}{\ell_\Pl}
\newcommand{\tpl}{t_\Pl}

\newcommand{\Mpl}{M_\Pl}

\newcommand{\mhyp}{\,\mbox{--}\,}

\newcommand{\au}[2]{#1.~#2}
\newcommand{\book}[5]{\emph{#1}, #2, #3, #4 (#5)}
\newcommand{\books}[4]{\emph{#1}, #2, #3 (#4)}
\newcommand{\oarX}[1]{\href{http://arxiv.org/abs/#1}{{\ttfamily\cob arXiv:#1}}}
\newcommand{\arX}[1]{\href{http://arxiv.org/abs/#1}{{\ttfamily\cob arXiv:#1}}}
\newcommand{\doin}[6]{\href{http://dx.doi.org/#1}{{\cob {\it #2} {\bf #3 #4} (#6) #5}}}
\newcommand{\doinn}[5]{\href{http://dx.doi.org/#1}{{\cob {\it #2} {\bf #3} (#5) #4}}}
\newcommand{\doij}[5]{\href{http://dx.doi.org/#1}{{\cob {\it #2} {\bf #3} (#5) #4}}}

\newcommand{\ndoinn}[5]{\href{#1}{{\cob {\it #2} {\bf #3} (#5) #4}}}

\newcommand{\tia}[1]{\textit{#1},}
\newcommand{\tiaq}[1]{\textit{#1},}

\newcommand{\boxd}[1]{\boxed{\phantom{\Biggl(}#1\phantom{\Biggl)}}}

\renewcommand{\leq}{\leqslant}
\renewcommand{\geq}{\geqslant}
\newcommand{\Eq}[1]{(\ref{#1})}
\newcommand{\Eqq}[1]{eq.~(\ref{#1})}
\newcommand{\Eqqs}[1]{eqs.~(\ref{#1})}
\def\rme{e}
\def\rmd{d}
\def\rmi{i}

\def\a{\alpha}
\def\b{\beta}
\def\de{\delta}

\def\g{\gamma}
\def\la{\lambda}
\def\k{\kappa}
\def\e{\epsilon}

\def\Om{\Omega}
\def\om{\omega}
\def\G{\Gamma}
\def\t{\tau}
\def\s{\sigma}

\def\N{\nabla}
\def\B{\Box}
\def\H{{\rm H}}

\def\Lst{\Lambda_*}
\def\Lz{\Lambda_*}
\def\Est{\Lambda_{\rm hd}}
\def\lst{\ell_*}

\def\Hes{{\bf\Delta}}

\def\cB{\mathcal{B}}

\def\cL{\mathcal{L}}

\def\cN{\mathcal{N}}
\def\cO{\mathcal{O}}
\def\cP{\mathcal{P}}

\def\cR{\mathcal{R}}

\def\cT{\mathcal{T}}

\def\p{\partial}

\def\cob{\color{blue}}

\begin{document}

\renewcommand{\thefootnote}{\fnsymbol{footnote}}

\title{Testing quantum gravity with primordial gravitational waves}

\author[a,*]{Gianluca Calcagni\,\orcidlink{0000-0003-2631-4588}\note{Corresponding author.}}
\emailAdd{g.calcagni@csic.es}
\affiliation[a]{Instituto de Estructura de la Materia, CSIC, Serrano 121, 28006 Madrid, Spain}

\author[b,c]{and Leonardo Modesto\,\orcidlink{0000-0003-2783-8797}}
\emailAdd{leonardo.modesto@unica.it}
\affiliation[b]{Dipartimento di Fisica, Universit\`a di Cagliari, Cittadella Universitaria, 09042 Monserrato, Italy}
\affiliation[c]{Department of Physics, Southern University of Science and Technology, Shenzhen 518055, China}


\abstract{We propose a testable alternative to inflation directly built in a very general class of ultraviolet complete theories of quantum gravity enjoying Weyl invariance. After the latter is spontaneously broken, logarithmic quantum corrections to the action make both the primordial tensor spectrum (from graviton fluctuations) and the scalar spectrum (from thermal fluctuations) quasi scale invariant. We predict a scalar spectral index $n_{\rm s}$ which only depends on two parameters and is consistent with observations, a tensor index $n_{\rm t} =1-n_{\rm s}>0$, and, if the fundamental energy scale of the theory $\Lambda_*=M_{\textrm{Pl}}$ is of order of the Planck mass, a tensor-to-scalar ratio $r_{0.05}\approx 0.01$ detectable by BICEP Array and LiteBIRD in the immediate or near future, respectively, and a blue-tilted stochastic gravitational-wave background observable by DECIGO in the further future. From the observational constraint $r_{0.05}<0.036$, we also find the lower bound $\Lambda_*> 8.5\times 10^{10}\,{\rm GeV}$, much stronger than any previous one for this class of theories.}


\maketitle
\renewcommand{\thefootnote}{\arabic{footnote}}


\section{Introduction}

The early universe is an ideal laboratory where to test theories beyond Einstein gravity and the Standard Model of quantum interactions. Microscopic scales can inflate to cosmological ones in the blink of an eye; extreme curvature regimes can dominate near the big bang or a big bounce; exotic particles can show their invisible presence through the cosmic evolution of the large-scale structures; and so on. Unfortunately, in general, one has to strike a compromise between theoretical control and predictivity, and the more rigorous and assumption-free is the theory, the less it makes contact with observations, and vice versa. For example, quantum gravity has made many of its advances to phenomenology in the context of inflation (see, e.g., \cite{Calcagni:2017sdq} for an overview) but, until now, it has been challenging to overcome theoretical difficulties and, at the same time, extract predictions that could be falsified in the near future.

In its purest form as a single scalar field, the inflationary mechanism requires a considerable amount of engineering, from the choice of potential to the embedding in a particle-physics or quantum-gravity framework. It is significant that the most successful scenario of early-universe acceleration to date \cite{Akrami:2018odb}, Starobinsky inflation \cite{Starobinsky:1980te}, is actually a model of gravity where the scalar degree of freedom is hidden, and that such model is not ultraviolet (UV) complete \emph{per se} but requires an embedding in a larger theory of quantum gravity, such as those attempted in supergravity \cite{Cecotti:1987sa,Ellis:2013xoa,Kallosh:2013lkr}, nonlocal gravity \cite{Briscese:2012ys,Briscese:2013lna,Koshelev:2016xqb,Koshelev:2017tvv,Calcagni:2020tvw}, or string theory \cite{Blumenhagen:2015qda}. Considering also that string theory is facing a swampland crisis \cite{Ooguri:2006in,Obied:2018sgi,Palti:2019pca} and it seems difficult to realize single-field inflation therein, the drive to look for alternatives might be felt even more urgently by some.

In this paper, instead of trying to get inflation from a fundamental theory, we propose a candidate with fundamental nonlocality which contains, basically for free, the ingredients needed to solve the singularity, horizon and flatness problems of the hot big bang \cite{Modesto:2022asj} and to explain the anisotropies of the cosmic microwave background (CMB), all without inflation. These achievements are due to an anomaly-free symmetry, Weyl invariance, and its spontaneous breaking; all the other details of the theory are unessential. As a phenomenological mechanism, Weyl invariance as an alternative to inflation is not new \cite{Antoniadis:1996dj,Antoniadis:2011ib,Amelino-Camelia:2013gna,Amelino-Camelia:2015dqa} but, to date, it was not known how to make it work in a fundamental theory due to the challenge of obtaining UV finiteness. The novelty here lies, on one hand, in the fact that we provide a concrete setting where quantum Weyl invariance is realized and, on the other hand, in the extraction of rigid predictions on the tensor sector which can be not only verified, but even falsified by present- and next-generation gravitational-wave (GW) observatories. Such rigidity comes from the fact that the cosmological model discussed below is derived \emph{directly} from the full theory under assumptions that can hardly be relaxed without spoiling the self-consistency of the theory. 

In particular, we find a lower bound on the tensor-to-scalar ratio $r$ at CMB scales observable by BICEP-Array, as well as a novel $O(10^{10}\,{\rm GeV})$ lower bound on the fundamental energy scale $\Lst$ of the theory from the \textsc{Planck}-BICEP-Keck observational upper bound on $r$. For these reasons, this is arguably one of the sharpest instances, compared with few others \cite{Calcagni:2020tvw,BC1,Addazi:2021xuf,LISACosmologyWorkingGroup:2022jok,LISA:2022kgy}, where the observation of the GW universe is opening a robust window on microphysics.

The paper is organized as follows. In section \ref{sett}, we briefly recall the setting of nonlocal quantum gravity, the transition from a Planckian Weyl-invariant phase to a post-Planckian non-Weyl one and the resolution of the hot big-bang problems detailed in the companion paper \cite{Modesto:2022asj}. In section \ref{Perturbations}, we compute the primordial tensor and scalar spectra generated during the post-Planckian phase, their indices, the tensor-to-scalar ratio and the GW background, showing that the theory yields characteristic predictions in a wide range of frequencies that, once combined, can be discriminated from inflationary models. Conclusions are in section \ref{conclu}. Appendices contain technical material and calculations.


\section{Summary of the setting}\label{sett}

A theory of gravity nonminimally coupled with matter was introduced in \cite{Modesto:2021ief,Modesto:2021okr,Calcagni:2023goc}. Thanks to the presence of certain nonlocal operators (i.e., terms with infinitely many derivatives), the quantum properties of this proposal are very similar to those of a more general class of nonlocal theories which are asymptotically local \cite{Kra87,Kuz89,Tomboulis:1997gg,Modesto:2011kw,BGKM,MoRa1,MoRa2,Dona:2015tra,GiMo,Modesto:2021soh} and include perturbative unitarity (absence of classical instabilities or quantum ghost modes), same nonlinear stability properties as in Einstein gravity, same tree-level scattering amplitudes, same macro-causality properties, and super-renormalizability or finiteness at the quantum level. In four dimensions, the action is
\be
S = \int \rmd^4 x \sqrt{|g|} \left[\cL_{\rm loc} +E_i \, F^{ij}(\Hes) \, E_j  + V(E_i) \right],\label{action} \qquad \cL_{\rm loc} = \frac{\Mpl^2}{2}\,R[g]+\cL_{\rm m}+\cL^{\rm loc}_2\,,
\ee
where $\Hes_{ki} \coloneqq \de^2 S_{\rm loc}/\de\Phi_k\de\Phi_i$ is the local Hessian for all gravitational and matter fields $\Phi_i$, $E_i$ give the local equations of motion $E_i=0$, and $V(E_i)$ is a collection of local operators at least cubic in $E$ that secure the finiteness of the theory in even dimensions \cite{MoRa1,Calcagni:2023goc}. The local extremals with respect to the metric are $E_{\mu\nu} \coloneqq(\Mpl^2 G_{\mu\nu} - T_{\mu\nu})/2$, where $\Mpl^2\coloneqq(8\pi G)^{-1}$ is the reduced Planck mass, $G_{\mu\nu}=R-g_{\mu\nu}R/2$, and $T_{\mu\nu}$ is the energy-momentum tensor derived from the matter Lagrangian $\cL_{\rm m}$. The term $\cL^{\rm loc}_2$ will be explained shortly. The nonlocal equations of motions of the theory \Eq{action} in the gravitational sector are
\be
\big[\rme^{\bar\H(\Hes)}\big]_{\mu\nu}^{\s\t} \, E_{\s\t} +O(E_{\mu\nu}^2)=0\,,\label{LEOM0}
\ee
where $\bar{\H}(\Hes) \coloneqq \H(\Hes)-\H(0)$ is an entire analytic function of the Hessian entering the definition of the form factor $F$:
\bs\label{fofa}\ba
\hspace{-1.1cm}&&2 \Hes F(\Hes) \equiv \rme^{\bar\H(\Hes)} - 1=\rme^{\g_\textsc{e} + \G[0,p(z)]} \,  p(z)-1\,,\label{FF}\\
\hspace{-1.1cm}&&\H(\Hes) =  \g_\textsc{e} + \G[0, p(z)] + \ln p(z) \,,\qquad z=\frac{4\Hes}{\Mpl^2\Lst^2},\label{H}
\ea
where $\Lst$ is an energy scale, $\g_\textsc{e}\approx 0.577$ is the Euler--Mascheroni constant, $\G$ is the upper incomplete gamma function, and $p(z)$ is the quartic polynomial
\be
p(z) =b_0-b\,z+z^4\,, \label{Poly}
\ee\es
where $b>0$ (this sign choice will be dictated by \Eq{nS}). Note that, with the choice \Eq{Poly}, $\H(0)=0$ when $b_0=0$ ($\bar\H=\H$); later on we will drop $b_0$ altogether, since it is negligible in the energy regimes we will consider. The form factor \Eq{fofa} simplifies previous proposals \cite{Kuz89,Modesto:2011kw} and guarantees a well-defined classical problem of initial conditions \cite{CMN3}, the absence of extra poles in the propagators, a unique Lorentzian limit \cite{Efimov:1967dpd,Pius:2016jsl,Briscese:2018oyx}, unitarity, and power-counting super-renormalizability in four dimensions \cite{Calcagni:2023goc} (which is essentially given by the highest-order operators $\cR^2\to h \B^2 h$ of Stelle gravity \cite{Ste77}, where $h$ is the graviton and $\cR=R_{\mu\nu},R$). In fact, in the UV the form factor is asymptotically polynomial:
\be
\rme^{\H(z)}\stackrel{z\gg 1}{\simeq} \rme^{\g_\textsc{e}} p(z)\,,\label{UVlimit}
\ee
and one can perform the usual power counting of divergences in Feynman diagrams. Note for later that, if $b \gtrsim O(1)$, then there exists an intermediate UV regime where the linear term dominates over the quartic one.

Finally, the metric in \Eqq{action} is defined as $g_{\mu\nu}\coloneqq (\sqrt{2}\phi/\Mpl)^2 \, \hat{g}_{\mu\nu}$, where $\phi$ is the dilaton field and the $\sqrt{2}$ factor is for convenience. All matter fields are rescaled similarly, so that the theory is classically invariant under Weyl transformations\footnote{The local  Weyl symmetry considered in this paper is expressed by the transformation \Eq{WI} with \emph{arbitrary} conformal factor $\Om(x)$, which acts on the fields of the theory and leaves the action invariant. One should not confuse this with the global conformal symmetry of a spacetime, which leaves the line element invariant. The latter symmetry is expressed by the conformal coordinate transformations of the conformal group of such spacetime. For instance, the group associated with conformal invariance of $D=4$ Minkowski spacetime is ${\rm SO}(2,4)$, the extension of the Poincar\'e group comprising not only translations, rotations and Lorentz boosts, but also coordinate dilatations and the so-called special conformal transformations ${x'}^\mu=(x^\mu-b^\mu x^2)/(1-2b_\mu x^\mu+b_\mu b^\mu x^2)$ \cite{DiFrancesco:1997nk,Sch08}. Acting with these coordinate transformations leaves the line element $\rmd s^2=\eta_{\mu\nu}\rmd x^\mu \rmd x^\nu=g'_{\mu\nu}\rmd {x'}^\mu \rmd {x'}^\nu$ invariant and the metric $g_{\mu\nu}'=\Omega^2(x)\eta_{\mu\nu}$ is rescaled by a conformal factor $\Omega(x)=1-2b_\mu x^\mu+b_\mu b^\mu x^2$, which is \emph{not arbitrary} but is fixed by the four parameters $b^\mu$ of the special conformal transformations.}
\be
\hat{g}^\prime_{\mu\nu} = \Om^2(x)\, \hat{g}_{\mu\nu} \, , \qquad \phi^\prime = \Om^{-1}(x)\, \phi \,,\qquad \dots\,. 
\label{WI}
\ee
It is well-known that the introduction of the dilaton in this way can be applied to \emph{any} theory and it obviously removes all the scales in the action, since all dimensionful couplings become numbers. This does not mean that any dynamical theory is Weyl invariant, since, in general, this symmetry is broken at the quantum level. Then, the choice
\be\label{Wesib}
\phi=\frac{\Mpl}{\sqrt{2}}
\ee
is simply a gauge fixing removing a redundant degree of freedom and all the procedure is trivial. However, in the present case the theory is also Weyl-invariant at the quantum level, since it is finite, all beta functions vanish, and the Weyl anomaly is zero \cite{Calcagni:2023goc}. Then, Weyl symmetry is preserved and gauge fixing (and, in particular, the choice \Eq{Wesib}) is tantamount to a spontaneous symmetry breaking (more on this below and in footnote 4 of \cite{Modesto:2022asj}).

As a consequence, the theory has no physical energy scales, since all couplings are rescaled by powers of the dilaton field $\phi\equiv\Mpl/\sqrt{2} + \varphi$. Still, one can define a UV and an infrared (IR) regime when comparing the energy component of the momentum and the scale $\Lst$ appearing in the form factors, which then become physical after Weyl symmetry is broken \cite{Modesto:2022asj}. At UV energy scales $E\coloneqq k^0\gg \Lst$, 
the scattering amplitudes are suppressed \cite{Jax}. At lower energy scales $E\ll\Lst$, the quantum Lagrangian acquires logarithmic corrections $\cL_{Q}$ just like Stelle gravity:
\be
\cL_{Q} = \b_R R \ln \left(\frac{-\B}{\Lst^2\de_0^2}\right) R + 
 \b_{\rm Ric} R_{\mu\nu}  \ln \left(\frac{-\B}{\Lst^2}\right) R^{\mu\nu}, \label{LQ}
\ee
where $\de_0$ is a constant to be determined by an explicit calculation we do not need here. At energy scales $E\sim\Mpl$, the theory is manifestly Weyl-invariant, there is no absolute notion of spacetime distance, and the correlation functions of fields with nonzero conformal weight are identically zero as a consequence of conformal symmetry \cite{DiFrancesco:1997nk}. In this \emph{Weyl phase}, the quantum corrections \Eq{LQ} always feature three or more fields and, therefore, do not enter perturbative two-point functions.

In order to recover the nonconformal world we observe, a symmetry breaking mechanism must be in place. In this theory, Weyl symmetry is broken spontaneously at scales around the Planck energy and the ensuing nonconformal phase begins already at early times. One way to implement this is to include a Weyl-invariant term $\cL^{\rm loc}_2$ in the local Lagrangian $\cL_{\rm loc}$ which provides a potential for the dilaton with nontrivial minima \cite{Modesto:2022asj}. A physically alluring possibility is $\cL^{\rm loc}_2$ to be the Weyl-invariant Higgs-dilaton potential proposed in \cite{Bars:2006dy,Bars:2013yba}:
\be\label{hidil}
\cL^{\rm loc}_2=\la (\mathfrak{h}^\dagger\mathfrak{h}-\a\phi^2)^2+\la'\phi^4\,,
\ee
where $\mathfrak{h}$ is the Higgs field. Notably, for $\la'\geq 0$ the dilaton and the Higgs field share the set of minima $\mathfrak{h}^\dagger\mathfrak{h}=\a\phi^2$ such that the expectation value of the former fixes the one of the latter and vice versa. In particular, by gauge fixing (recall that dilaton perturbations are gauge-dependent, so that a choice of constant $\phi$ breaks Weyl invariance even for a potential with flat directions), at one of these minima the dilaton acquires the expectation value $\phi=\Mpl/\sqrt{2}$, whose value is determined empirically by the measurement of Newton's constant. Then, Weyl symmetry is broken and the system acquires dimensionful couplings which can no longer be rescaled arbitrarily by a Weyl transformation. When $\varphi \ll \Mpl$ and the dilaton sits in its expectation value, we enter a sub-Planckian \emph{Higgs phase} where a background is selected and one can consider cosmological perturbations thereon. At the beginning of this phase, around energies $E\sim \Est$ with
\be\label{Estar}
\Est\coloneqq   \sqrt{\frac{{\rm max}(1,b_0)}{b}}\,\Lst
\ee
(actually, it will turn out that we can drop $b_0$ so that $\Est=\Lst/\sqrt{b}$), the dynamics is still nonlocal (we are not truncating the form factor at low momenta\footnote{Truncation of nonlocal operators to a finite order are notoriously delicate and can introduce artificial solutions and degrees of freedom \cite{Eliezer:1989cr,Moeller:2002vx,Calcagni:2007ru}.}) but the asymptotic limit of the form factor is such that the gravitational sector becomes local and quadratic in the Ricci tensor \cite{Modesto:2022asj}:
\ba
\cL_{\Est} &\simeq& \frac{\Mpl^2}{2} R + \cL_{\rm m} + F(0) E_{\mu\nu} E^{\mu\nu}+ \cL_{Q}\nn
&\simeq&\frac{\Mpl^2}{2} R+\cB R_{\mu\nu} R^{\mu\nu}+\cL_{\rm m}+ \cL_{Q}\,,\qquad \cB=-\frac{\rme^{\g_\textsc{e}-\H(0)}b}{2}\frac{\Mpl^2}{\Lz^2}\,,\label{modeB}
\ea
in $D=4$ dimensions and dropping sub-leading $T_{\mu\nu}T^{\mu\nu}$ terms (the expression of $E_{\mu\nu}E^{\mu\nu}$ with matter is \cite[eq.\ (A.5)]{Modesto:2022asj}). The quantum corrections \Eq{LQ} contribute to the graviton propagator. At yet lower energies, $\cL\sim R$ and Einstein gravity is recovered.

In \cite{Modesto:2022asj}, we started the description of the cosmology stemming from this theory. To recapitulate, there we concentrated on the background, leaving perturbations to the present paper. Weyl invariance provides an extremely simple, natural, and model-independent solution to the problems of hot-big-bang cosmology. All these issues are addressed in the Weyl phase of the theory, while the quasi scale-invariant primordial temperature spectrum arises in the Higgs phase. For details, see \cite[sections 3.3 and 4]{Modesto:2022asj}.

Singularities in a theory should be resolved at very short scales and, in the case of the big bang, very short times, in the conditions where and when such singularities would form. When a resolution mechanism is provided by a symmetry enforced at such times and distances, then it works even if the symmetry is broken at larger scales or later times. In the present context, the big-bang problem is solved because the Borde--Guth--Vilenkin theorem, establishing geodesic incompleteness in a very general class of gravitational theories \cite{BGV}, does not apply here. In fact, the average expansion condition at the core of the theorem is not a Weyl-invariant statement and cannot be met in a theory where the dynamics is Weyl-invariant. Vice versa, a Weyl transformation can always make an expanding background static or even contracting and any singular metric can be rescaled into a regular one. In particular, the Friedmann--Lema\^{i}tre--Robertson--Walker (FLRW) metric can be mapped into asymptotically flat metrics. After Weyl invariance is broken, the hypotheses of the Borde--Guth--Vilenkin theorem are met but, obviously, they cannot be extrapolated at arbitrarily early times beyond the point where Weyl invariance is restored.

The flatness problem is dispensed with by pure symmetry arguments. Weyl invariance forces all regions of the Universe to be in causal contact, which eventually leads to select homogeneous and isotropic metrics (i.e., Minkowski and FLRW with intrinsic curvature $\textsc{k}=0,\pm1$) at the onset of the Higgs phase. The well-known observations that FLRW spacetime with any $\textsc{k}=0,\pm1$ is conformally equivalent to Minkowski \cite{Narlikar:1986kr,Iihoshi:2007uz,Ibison:2007dv} in a Diff invariant theory and that, moreover, all FLRW metrics are also physically equivalent to the FLRW spacetime with zero intrinsic curvature in a Weyl$\times$Diff invariant theory further reduce the possibilities to Minkowski and FLRW with $\textsc{k}=0$.

The horizon problem does not arise because distances do not have any absolute meaning in the Weyl phase. At very high energies and early times, Weyl symmetry forces spacetime to be infinite and wholly causally connected. Thus, there is no horizon problem from the outset. After Weyl symmetry is broken, the radius $H^{-1}$ of the Hubble horizon acquires physical meaning and, as shown in \cite[section 5]{Modesto:2022asj}, it is equal to or larger than the radius $H^{-1}_{\rm Ein}$ of FLRW in Einstein gravity and tends asymptotically to it, $H^{-1}\to H^{-1}_{\rm Ein}$. The standard evolution of the universe is recovered already during radiation domination when $H^{-1}\simeq H^{-1}_{\rm Ein}$.\footnote{Both in the standard cosmological model and in the conformal model under study, our large observable universe originated from an initially small causal patch of size $\sim H^{-1}$ that later evolved for 13-14 billion years of radiation-matter-$\Lambda$-dominated expansion. The main difference is how one explains the fact that also the regions outside this patch, that later entered the horizon during the history of the universe, were already in causal contact with one another and had the same thermal properties of the patch. In the standard cosmological model, this is achieved with inflation.} Note that the mechanism super-selecting the background at the end of the Weyl-invariant phase is based on symmetry and is therefore time-independent, in the sense of being enforced during the whole duration of the Weyl phase. Thus, it makes the background conditions at the end of such phase (final conditions) and at the beginning of the Higgs phase (initial conditions) stable against perturbations, which can be rendered arbitrarily small by a Weyl transformation \cite[section 4.1]{Modesto:2022asj}. Stability of the background is maintained also after Weyl symmetry is broken. In fact, since radiation-dominated flat FLRW is a solution of Einstein gravity, then it is also a solution of the classical nonlocal theory \cite{Modesto:2022asj,Briscese:2019rii} with the same stability properties \cite{Briscese:2019rii}. By construction, the quantum corrections $E^Q$ in \Eqq{linear0} do not disrupt these properties.

The monopole problem is avoided because in asymptotically local quantum gravity all the couplings run to zero in the UV, similarly to the scenario of \cite{Robinson:2005fj}. In the Weyl phase, there are no causally disconnected domains where the Higgs field of a grand unification theory could take different expectation values.

Finally, the trans-Planckian problem does not occur because, while in standard inflation primordial perturbations with sub-Planckian wavelengths eventually become observable, here the Weyl phase does not admit a comparison of physical lengths with any given scale. From the point of view of quantum field theory, the higher-derivative operators $E_i (\B/\Mpl^2)^n E^i$, important at or above the Planck energy, are quadratic in $E_i$ and do not contribute to the linearized nonlocal equations of motion
\be
\rme^{\bar\H(\Hes)^{(0)}} E^{(1)}_i+E^Q_i+O(E^2_i) = 0 \, , \label{linear0}
\ee
where the superscripts denote, respectively, the background and the first order in field perturbations and $E^Q$ is the contribution from the quantum corrections \Eq{LQ}. In particular, all background solutions of the Einstein's equations $E_{\mu\nu}=0$ (including FLRW metrics) are exact solutions of the theory \Eq{action} at the classical level and approximated solutions at the quantum level (by the construction in appendix \ref{appeB3}, quantum corrections are sub-dominant). Examples of backgrounds, meaningful only after spontaneous symmetry breaking, that are exactly flat and geodesically complete are given in \cite{Modesto:2022asj}.


\section{Primordial spectra}\label{Perturbations}

While the Weyl phase can address the classic problems of the hot-big-bang model, it does not explain the origin of the primordial spectra and their almost scale invariance. It turns out that these are generated during the post-Planckian phase, where Weyl invariance is spontaneously broken. The one-loop quantum corrections to the classical action are directly responsible for the small deviation from Harrison--Zel'dovich spectra.

In this section, after describing some basic approximations needed for physical and technical reasons, we derive the quasi-scale-invariant spectrum of primordial tensor and scalar perturbations. 
 To this aim, we need the graviton propagator for the theory \Eq{action} with form factor \Eq{fofa}. The tree-level expression can be found in appendix \ref{appeB2}, while the graviton propagator from the one-loop quantum effective Lagrangian \Eq{modeB} is calculated in appendix \ref{appeB3}. 
 
We do not consider vector perturbations because, in the full nonlocal theory, they decay in time for any expanding power-law background, not necessarily inflationary, as one can see by applying perturbation theory on the Einstein-gravity solutions of $E_{\mu\nu}=0$ \cite{Durrer:2004fx}. This result is valid in the full theory in the nonlocal trans-Planckian phase, where $E_{\mu\nu}=0$ are exact classical solutions, but is not altered even at later times when the dynamics approaches the quadratic regime \Eq{modeB}, where $\cB<0$. The background and perturbative cosmology of this Lagrangian \cite{Whitt:1984pd} have been studied in detail \cite{Noh:1996da,Hwang:1997kf,Noh:1998hd,Yun:2020jxx,Yun:2022xce,Park:2022lgf}. In particular, from eqs.\ (15), (25) and (26) of \cite{Hwang:1997kf}, one can see that the vorticity velocity $v_\om$ is constant in a radiation-dominated universe. The equation of motion (25) for the difference $\Psi \coloneqq v_\s - v_\om$ between shear and vorticity velocity shows that $\Psi$ decays as $\sim a^{-2}$ as in Einstein gravity ($\cB=0$), since the $O(\cB)$ terms are further suppressed in time.


\subsection{Flat spacetime approximation}\label{PerturbationsA}

According to the discussion in \cite[section 3.2]{Modesto:2022asj} 
 and the results derived in appendix \ref{appeB3}, the computations we are going to make in this section rely on two assumptions for the energy regime in which we will consider the theory \Eq{action}:
\begin{enumerate}[label=(\roman*)]
	\item Quadratic-gravity or higher-derivative (hd) regime \cite{Modesto:2022asj}
	\be\label{interm}
	\boxd{\Est\lesssim E\lesssim\Lst\,.}
	\ee
	In this range, the theory has the local limit \Eq{modeB} which captures its main UV properties.\footnote{More generally, what follows holds also for Stelle gravity with a nonzero coefficient of the $R^2$ operator.} The reason why one should consider the quadratic-curvature regime of the theory is not just convenience (the theory in the fully nonlocal regime may be very difficult to handle cosmologically), but also empirical. We know that early-universe spectra are quasi-scale-invariant. Then, if they are generated by curvature (as they should in this model, since we do not have matter fields sustaining inflation), they must proceed from operators of the same dimensionality of spacetime. Thus, in four dimensions these operators are quadratic curvature terms, so that the dimensional scaling of the graviton two-point function in this regime is $\sim 1/k^4$. Thus, if the theory can account for the primordial cosmological spectra, it must do so in the Stelle-like regime. Quantum corrections naturally break exact scale invariance, as described below.
	\item Perturbative (one-loop) approximation \Eq{z01}, corresponding to $E \gtrsim \Lst/10$. It guarantees that one-loop quantum corrections in the propagator are subdominant with respect to the classical (tree-level) part consistently with the quantum field theory perturbative expansion. This condition also permits some analytic simplifications (\Eqqs{1pq1} and \Eq{1pq2}), but its main reason is to ensure that we only explore corners of the parameter space where perturbative quantum gravity does not break up.
\end{enumerate}
Therefore, if ${\rm max}(1,b_0)/b<O(10^{-2})$, 
 then $\Est<\Lst/10$ and the conditions (i) and (ii) imply that $\Lst/10\lesssim E\lesssim\Lst$, which boils down to pushing the energy to the lower limit:
\be
\boxd{E \simeq \frac{\Lst}{10}\,.}
\label{energyRegime}
\ee
If we make the identification $\Lst = \Mpl$, then the regime \Eq{energyRegime} falls within the post-Planckian or Higgs phase where Weyl invariance is spontaneously broken. 

In \cite[section 3.1]{Modesto:2022asj}, we argued that $\Lst\leq\Mpl$ only using Weyl invariance. Here, we end up with a weaker statement using \Eq{energyRegime} and the fact that the energy of particles propagating in the Higgs phase must be smaller than the Planck mass, $E < \Mpl$:
\be
\Lst < 10 \, \Mpl\,, \qquad \lst > \frac{\lp}{10}\,.
\label{LambdaMp2}
\ee
where $\lst\coloneqq \Lst^{-1}$ and $\lp\coloneqq\Mpl^{-1}$. 

In order to compute the primordial scalar and tensor spectrum, we first show that we can approximate the asymptotic FLRW metric with the Minkowski one. Then, we can evaluate the correlation functions in flat spacetime and avoid the complications of quantum field theory on curved spaces. This approximation is not a novelty in certain scenarios of conformal cosmology \cite{Antoniadis:1996dj,Antoniadis:2011ib} and cosmologies from quantum gravity \cite{Bonanno:2001xi,Bonanno:2007wg} where the correct primordial spectra are generated subhorizon. The condition that has to be fulfilled is to consider perturbations within the Hubble horizon,
\be\label{condition}
\la = \frac{1}{E} \ll R_H=H^{-1}\,,
\ee
where $\la$ is the wavelength of the perturbation. In the following, we analyze the timeline of the cosmological model and the hierarchy that its characteristic time scales must satisfy to comply with \Eq{condition}.

There are three important moments in the evolution of the Universe: the instant $t_{\Lst}$ when $\la=\lst$ and the Universe exits the Weyl phase, the time (or, more precisely, a narrow range of time scales) $t_\la$ when $\la\simeq 10\lst$ and the wavelength has the same scale of validity of the Stelle regime (see \Eqq{energyRegime}), and the instant $t_{\Est}$ when $\la=\Est^{-1}$ before which the theory can be approximated by the local quadratic action \Eq{modeB} and after which it starts to recover Einstein gravity. Apart from these, there is another time instant of interest $t_\star$ which signals the moment after which the cosmic expansion becomes almost indistinguishable from the one in Einstein gravity. As we will see shortly, the recovery of the background-independent Einstein action does not happen exactly at the same time of the recovery of the standard cosmic expansion, and $t_{\Est}\neq t_\star$. In fact, since the class of FLRW metrics is an exact solution of the full theory ($E_{\mu\nu}=0$, which solves \Eqq{LEOM0}), the standard cosmic expansion is expected to be reached some time before $t_{\Est}$, so that $t_{\Est}> t_\star$. This is indeed the case and $t_\star$ is determined as follows.

Recall that Weyl invariance in the trans-Planckian phase forces the background at the end of such phase to be either Minkowski or FLRW with intrinsic curvature $\textsc{k}=0$, whatever the initial conditions \cite{Modesto:2022asj}. When Weyl invariance is broken, perturbations around this background become physical and one can generate the cosmological spectra. Before that, however, we have to see how to connect the background at the beginning of the Higgs phase (Minkowski or flat FLRW) with the one at later times, for instance by some of the interpolating exact solutions of \cite[section 5]{Modesto:2022asj}. In particular, to match observations, 
 we have to consider a metric being asymptotically FLRW at late times when Einstein gravity is recovered. The most natural setting in all scenarios alternative to inflation is a radiation-dominated Universe described by the scale factor
\be\label{raddo}
a(t) \simeq \sqrt{\frac{t}{t_0}}\,, 
\ee
which we will discuss at the beginning of section~\ref{ScalarPerturbations}. Note that this background is a phase-space attractor, i.e., it is future-stable against homogeneous perturbations.\footnote{We can apply the same analysis as in Einstein gravity, which we recall here. From the Friedmann equation $H^2=\rho/(3\Mpl^2)$ and the continuity equation $\dot\rho+3H(1+w)\rho=0$ for a perfect fluid with pressure $P$, energy density $\rho$, equation of state $P=w\rho$, and constant barotropic index $w$, one defines the phase-space variables $x\coloneqq H$ and $y\coloneqq \rho/\Mpl^2$, which obey the system $\dot x= -[(1+w)/2]y=-[3(1+w)/2]x^2$ and $\dot y= -3(1+w)xy$. Expanding $x(t)=x_0(t)+\de x(t)$ around the background $x_0(t)=2/[3(1+w)t]$ (scale factor $a(t)=(t/t_0)^{2/[3(1+w)]}$) and linearizing with respect to the homogeneous perturbation $\de x$, one finds $\dot \de x=-3(1+w)x_0\de x=-(2/t)\de x$, whose solution is $\de x(t)\propto t^{-2}$. This profile is suppressed as $t\to\infty$, regardless of the value of $w$. For radiation, $w=1/3$.}

\begin{figure*}
	\bc
	\includegraphics[width=\textwidth]{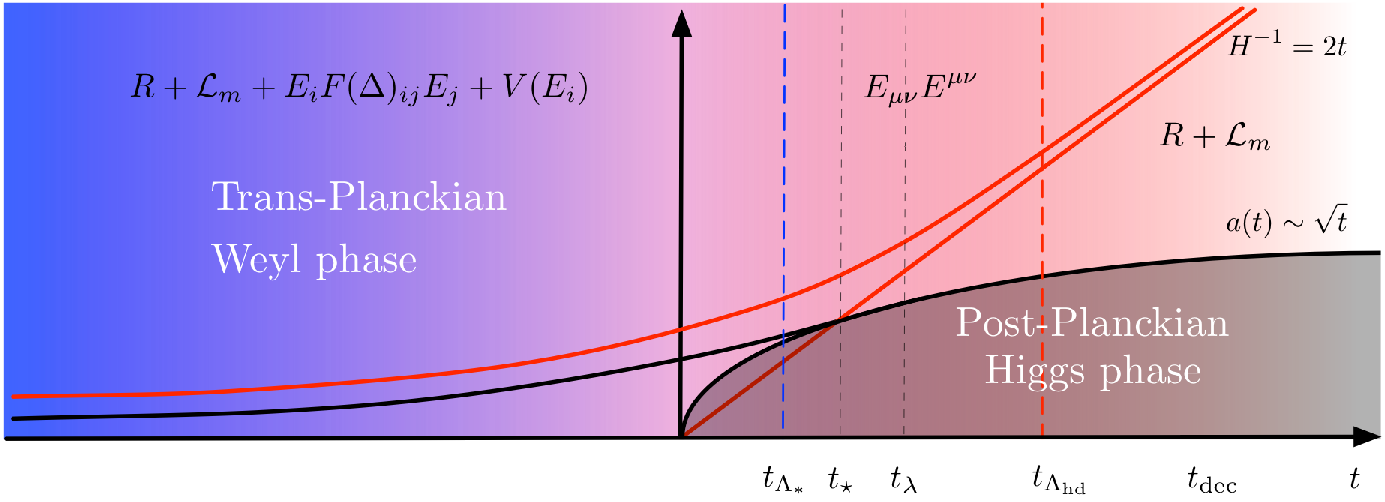}
	\ec
	\caption{The cosmological scenario proposed in this article. The black curve with support on the whole $t$ line is the scale factor $a = \sqrt{A + t/t_0 \, \rme^{\G (0,t/\tpl)}}$ \cite[eq.\ (5.9a) and figure 4a]{Modesto:2022asj}, which is chosen only for definiteness among any of the other solutions described in \cite{Modesto:2022asj} reducing to FLRW at late times. The black curve with support on the $t>0$ half-line is the FLRW radiation-dominated scale factor \Eq{raddo}. The red curve and red straight line are the Hubble radius associated with the above scale factors, respectively. For $t<t_{\Lst}$, the Universe is in the trans-Planckian Weyl phase described by the theory \Eq{action}. For $t \gtrsim t_{\Lst}$, the Weyl symmetry is spontaneously broken. In the epoch $t_{\Lst}< t \ll t_{\Est}$, the Universe is well described by quadratic gravity (Stelle's theory) plus quantum corrections. For $t>t_\star$, the background solution is indistinguishable from a radiation-dominated Universe in Einstein gravity, although the Einstein--Hilbert action is recovered only at later times $t>t_{\Est}$. In Einstein gravity, $t_\star$ would represent the time of horizon crossing of the perturbation ($k=aH$), \Eqq{tbar}, but notice that all regions of the Universe are in causal contact at any time in the plot for asymptotically local quantum gravity (the Hubble radius is always larger than the scale factor). $t_\lambda$ is when the perturbation is evaluated after the spontaneous breaking of the Weyl symmetry, consistently with the perturbative approximation \Eq{energyRegime}. The CMB is formed at the decoupling time $t_{\rm dec}$.}\label{fig1}
\end{figure*}

In Einstein gravity, a perturbation living in this background would cross the horizon at a time $t_\star$ such that $\la(t_\star) = R_H(t_\star)$. At the instant $t$, the wavelength is given by $\lambda = a(t) \,\la_{\rm com}$, 
 while $H = \dot{a}/a \simeq 1/(2t)$ and $R_H = H^{-1} \simeq 2 t$. Therefore, at $t=t_\star$ we have
\be
\sqrt{\frac{t_\star}{t_0}}\,\la_{\rm com}  \simeq 2 t_\star\qquad \Longrightarrow \qquad t_\star \simeq \frac{\la_{\rm com}^2}{4t_0}\,,
\label{tbar}
\ee
where $\la_{\rm com}=2\pi/|\bm{k}|$ is the comoving wavelength and $|\bm{k}|$ is the comoving wavenumber. In Weyl asymptotically local gravity, if the condition \Eq{condition} holds, then there is no horizon crossing for primordial perturbations produced inside the horizon and $t_\star$ has the physical interpretation advanced above, as one can see in figure~\ref{fig1}. At later points $t>t_\star$ in the evolution of the universe, the actual horizon $R_H>2t$ is calculated from the exact metric and the condition \Eq{condition} holds also beyond $t_\star$.

Let us now find the instants $t_{\Lst}$, $t_\la$, and $t_{\Est}$ and provide the chronological order of the events for which \Eqq{condition} is reliable. We take the asymptotic form \Eq{raddo} for the scale factor, which is sufficient to establish the hierarchy of time scales. The instant $t_{\Lst}$ is given by
\be
t_{\Lst} \! : \,\, \sqrt{\frac{t_{\Lst}}{t_0}}  \, \la_{\rm com}  \simeq \lst
\qquad \Longrightarrow \qquad
t_{\Lst} \simeq \frac{\lst^2 t_0}{\la_{\rm com}^2}= \frac{\lst^2}{4 t_\star}\,. 
\label{TLambda}
\ee
The instant $t_\lambda$ is given by
\be
t_\lambda \! : \,\,\sqrt{\frac{t_\lambda}{t_0}} \la_{\rm com} \simeq  10 \, \lst
\qquad \Longrightarrow \qquad
t_\lambda 
\simeq 10^2\,t_{\Lst}\,. 
\label{Tlambda}
\ee
The instant $t_{\Est}$ is given by
\ba
&&t_{\Est} \! : \,\,  \sqrt{\frac{t_{\Est}}{t_0}}\, \la_{\rm com} \simeq \sqrt{\frac{b}{{\rm max}(1,b_0)}} \,  \lst
\qquad \Longrightarrow \qquad t_{\Est}  
\simeq\frac{b}{{\rm max}(1,b_0)}\,t_{\Lst}\,. 
\label{TLambdaC}
\ea
Since $b/{\rm max}(1,b_0)\gg 1$, we have the hierarchy
\be\label{scalesInequa0}
t_{\Lst}\ll t_\la  < t_{\Est}\,,
\ee
which is the time analogue of the interval \Eq{interm}. 

The condition \Eq{condition} is satisfied in the regime \Eq{energyRegime} when
\ba
\lambda\simeq 10 \, \lst \ll  H^{-1}\simeq 2 t_\la \qquad
\Longrightarrow \qquad
t_\lambda \gg 5\lst \,.
\label{tcond}
\ea
Plugging \Eqq{tbar} into \Eq{tcond}, we find
\ben
t_\lambda \gg 5\lst \simeq 10 \sqrt{t_\star\, t_{\Lst}} \, ,
\een
in which we now replace $t_{\Lst} \simeq t_\lambda/10^2$:
\be\label{tstla}
t_\star  \ll t_\la \,. 
\ee
Since $t_\star$ marks an event posterior to the onset time of the Higgs phase, we also have $t_{\Lst}<t_\star$ which, combined with \Eq{scalesInequa0} and \Eq{tstla}, yields the interval
\be\label{scalesInequa}
\boxd{t_{\Lst}<t_\star\ll t_\la  < t_{\Est}\,.}
\ee
The cosmic evolution (not just the scale factor $a(t)$) becomes standard at times $t>t_{\Est}$, so that the time range of production of primordial perturbations in this cosmological model is roughly $t_{\Lst}<t< t_{\Est}$. 
The Weyl and the Higgs phase as well as the scale hierarchy \Eq{scalesInequa} and the local UV limit of the theory are depicted in figure~\ref{fig1}.

To summarize, according to the assumptions of (i) being in the quadratic-gravity regime and (ii) never allowing quantum corrections to dominate tree-level contributions, we end up with a cosmological model valid at energies \Eq{energyRegime}. This energy regime is only one order of magnitude smaller than the fundamental energy scale $\Lst$, hence it is much closer to the deep UV nonlocal regime than to the IR limit corresponding to Einstein gravity. In turn, this implies that there exists a wide region in the parameter space such that the wavelength of the perturbations giving rise to the observed spectra are well within the Hubble horizon, \Eqq{condition}.


\subsection{Tensor perturbations} \label{TensorPerturbations}

After the spontaneous breaking of Weyl invariance, distances acquire a meaning, correlation functions become nonvanishing, and one can consider metric perturbations. In this noninflationary early-Universe scenario, tensor perturbations are the subhorizon metric quantum fluctuations corresponding to the graviton, in a regime where one can ignore the curvature of the background and the cosmological expansion. In contrast, in standard inflation metric fluctuations are induced by the quantum fluctuations of a scalar field and the spectrum is calculated at subhorizon scales but evaluated at horizon crossing.

The tensor spectrum in the regime \Eq{energyRegime} is readily obtained from the graviton propagator \Eq{PQEA3} derived in appendix \ref{appeB3}. The two-point correlation function for transverse and traceless tensor modes in four dimensions reads \cite{Antoniadis:2011ib}
\ba
\hspace{-.7cm}G^{(2)}_{ijkl}(x,x^\prime)\!\!  &\coloneqq&\!\!  \langle h_{ij}(x) h_{kl} (x^\prime) \rangle\nn
\hspace{-.7cm}\!\!&\stackrel{\textrm{\tiny \Eq{PQEA3}}}{=}&\!\! -{\rm C} \int \frac{\rmd^4 k}{(2\pi)^4} \, \rme^{\rmi k\cdot (x - x^\prime)} \frac{P^{(2)}_{i j k l}(k)}{k^4 \left(\frac{k}{\Lst}\right)^{2\e_2}}\nn
\hspace{-.7cm}\!\!&=&\!\! -{\rm C}\Lst^{2\e_2}\,P^{(2)}_{i j k l}(\p_x) \!\! \int \!\! \frac{\rmd^4 k}{(2\pi)^4} \, \frac{\rme^{\rmi k\cdot (x - x^\prime)}}{k^{4 + 2 \e_2 }}\,, 
\label{hh}
\ea
where Latin indices run over spatial directions, the constants ${\rm C}$ and $\e_2$ are given by \Eqqs{PQEA3} and \Eq{epsilons}, respectively,
\be\label{CbRic}
{\rm C} \coloneqq \frac{4\Lz^2}{b \,\rme^{\tilde{\g}_\textsc{e}}\Mpl^2}
\, ,  \qquad \e_2 = - \frac{2\b_{\rm Ric} \Lz^2}{b \, \rme^{\tilde{\g}_\textsc{e}}\Mpl^2}\,,\qquad \tilde{\g}_\textsc{e} \coloneqq \g_\textsc{e}-\H(0) 
\ee
and $P^{(2)}_{i j k l}(k)$ are the spatial components of the spin-$2$ projector in \Eqq{projectors} \cite{Riv64,Bar65,VanNieuwenhuizen:1973fi}. The angular brackets in the first line of \Eqq{hh} are defined via the Lorentzian path integral and the final expression is also Lorentzian. To perform the integral explicitly, however, we now move to Euclidean momenta as described in \cite{Calcagni:2024xku}. We recall that the two-point correlation function in $D$-dimensional Euclidean momentum space is obtained through the Fourier transform
\bs\label{xTOk}\ba
\hspace{-1cm}(x^2)^{-s} &=& B_D(2s) \int \frac{\rmd^D k}{(2\pi)^D} \rme^{\rmi k \cdot x} (k^2)^{s -\frac{D}{2}} \,,\\
\hspace{-1cm}B_D(2 s) &\coloneqq& \frac{2^{D-2s} \pi^{\frac{D}{2}}\G\left(\frac{D}{2}-s\right)}{\G(s)}\,.\label{xTOkb}
\ea\es
Therefore, comparing \Eqqs{hh} and \Eq{xTOk} with $D=4$, we see that $s=-\e_2$, so that the propagator in position space reads
\be
G^{(2)}_{ijkl}(x,x^\prime) = -\frac{{\rm C}\Lst^{2\e_2}}{B_4(-2\e_2)}\,P^{(2)}_{i j k l}(\p_x) |x-x^\prime|^{2\e_2}\,.
\ee
For $x_0^\prime = x_0$ and using again \Eqq{xTOk} with $s=-\e_2$, but now in $D-1$ dimensions, 
\ban
G^{(2)}_{ijkl}(\bm{x}, \bm{x}^\prime) &=& -\frac{{\rm C}\Lst^{2\e_2}}{B_4(-2\e_2)} 
\, P^{(2)}_{i j k l}(\p_x) |\bm{x} - \bm{x}^\prime|^{2\e_2} \\
&=& -\frac{{\rm C}\Lst^{2\e_2} B_{3}(-2\e_2)}{B_4(-2\e_2)} P^{(2)}_{i j k l}(\partial_x) \!\!
\int \frac{\rmd^{3} \bm{k}}{(2\pi)^{3}} \frac{\rme^{\rmi \bm{k} \cdot (\bm{x}- \bm{x}^\prime)}}{(\bm{k}^2)^{\e_2 +\frac{3}{2}}}\,. 
\ean
From \Eqq{xTOkb}, we can compute the ratio $B_{3}(-2\e_2)/B_4(-2\e_2)$:
\be
\frac{B_{D-1}(2s)}{B_D(2s)} 
=\frac{1}{2\sqrt{\pi}}\frac{\G\left(\frac{D-1}{2}\right)}{\G\left(\frac{D}{2}\right)}
\stackrel{D=4}{=} \frac{1}{4} \, .
\ee
Finally, according to the definitions in appendix \ref{appeD}, 
\ba
G^{(2)}_{ijkl}(r) &=& -\frac{{\rm C}\Lst^{2\e_2}}{4}P^{(2)}_{i j k l}(\p_x) \!\!\int \frac{\rmd^3 \bm{k}}{(2\pi)^3}\frac{\rme^{\rmi \bm{k} \cdot \bm{r}}}{k^{2\e_2+3}}\nn
&\stackrel{\textrm{\tiny \Eq{PGk}}}{=}& -P^{(2)}_{i j k l}(\p_x) \!\!\int_0^{+\infty}\rmd k\,\frac{k^2}{2\pi^2}\,P_h(k)\,\frac{\sin(kr)}{kr}\nn
&\stackrel{\textrm{\tiny \Eq{psipsiDe}}}{=}& -P^{(2)}_{i j k l}(\p_x) \!\!\int_0^{+\infty}\frac{\rmd k}{k}\,\Delta^2_h(k)\,\frac{\sin(kr)}{kr}\,,\label{kTOxd2}
\ea
where $\bm{r}=\bm{x}-\bm{x}^\prime$, $r=|\bm{r}|$, $k=|\bm{k}|$, and
\be
P_h(k)=\frac{{\rm C}\Lst^{2\e_2}}{4\,k^{2\e_2+3}} \,, \qquad \Delta^2_h(k)= \frac{{\rm C}}{8\pi^2}\left(\frac{\Lst}{k}\right)^{2\e_2}. 
\ee 
Taking into account the two polarization modes of the graviton, the final result for the tensor power spectrum $\cP_{\rm t}$ is
\be
\cP_{\rm t}(k) \coloneqq 2\Delta^2_h(k) =\frac{4\Lz^2}{b \,(2\pi)^2\,\rme^{\tilde\g_\textsc{e}}\,\Mpl^2}\left(\frac{k}{\Lst}\right)^{-2\e_2}.
\ee
Therefore, using the definition \Eq{epsilon4}, the tensor spectrum reads
\be
\boxd{\cP_{\rm t}(k) = \frac{n_{\rm t}}{(2\pi)^2 \,\b_{\rm Ric}}\left(\lst k\right)^{n_{\rm t}}\,,}
\label{DeltaTnT}
\ee
where the tensor spectral index is
\be
\boxd{n_{\rm t}\coloneqq \frac{\rmd\ln\cP_{\rm t}}{\rmd\ln k}=-2\e_2 =\frac{4\b_{\rm Ric}}{b\,\rme^{\tilde{\g}_\textsc{e}}}\frac{\Lz^2}{\Mpl^2}\,,} 
\label{nT}
\ee
which is positive because $\b_{\rm Ric} > 0$. 

Before moving to scalar perturbations, let us comment on how the quantum tensor modes turn classical. In the standard cosmological scenario, it is assumed that decoherence from a quantum field $\hat v_k$ to a classical field configuration $v_k$ takes place thanks to the presence of the cosmic horizon $H^{-1}$, which plays the role of a sort of ``measurement'' act forcing the quantum system into one of its eigenstates (see, e.g., \cite[sections 5.6.1 and 5.6.2]{Calcagni:2017sdq} and references therein). At the level of linear perturbations, the evolution across the horizon is governed by the Mukhanov--Sasaki equation for $\hat v_k$, which corresponds to a quantum ``squeezed'' state $|v_k\rangle$ undergoing a unitary time evolution \cite{Guth:1985ya,Grishchuk:1990bj,Grishchuk:1992tw,Albrecht:1992kf,Polarski:1995jg}. Then, the quantum-to-classical transition $\hat v_k\to v_k$ consists in the suppression of the effects of noncommutativity between the field $\hat v_k$ and its conjugate momentum $\hat\pi_v$ (for instance, the expectation values $\langle \hat v_k \hat\pi_v\rangle$ and $\langle \hat\pi_v \hat v_k\rangle$ on the squeezed state converge to each other). This understanding of cosmological decoherence is refined when nonlinear self-interactions of the perturbation field or interactions between the field and other degrees of freedom $\hat \psi_k$ are considered. Then, the pure squeezed state becomes a mixed one $|v_k,\psi_k\rangle$ where sub- and superhorizon modes interact with one another, time evolution becomes nonunitary, and small non-Gaussianities are generated. The density matrix $|v_k,\psi_k\rangle\langle v_k,\psi_k|$ acquires non-diagonal terms and each matrix entry is associated with a classical probability. Thus, actual decoherence consists in the decay of interference terms and it is in this sense that it has been studied for inflation in recent years \cite{Martin:2018zbe,Martin:2018lin,Gong:2019yyz,Brahma:2022yxu,Burgess:2022nwu,Burgess:2024eng,Brahma:2024yor}. Whether the same decoherence mechanism holds for the present non-inflationary nonlocal scenario remains to be seen but, as a starter, we obtain a positive answer at the linear level through the intuitive correspondence between the cross-horizon unitary evolution of the squeezed state and the behavior of the solutions of the Mukhanov--Sasaki equation for cosmological perturbations. In general, the dynamical equation for tensor modes in conformal time $\t=\int\rmd t/a(t)$ and in momentum space is
\be\label{musa}
\p_\t^2v_k+\left(c_{\rm t}^2k^2-\frac{\p_\t^2 z}{z}\right)v_k=0\,,
\ee
where $v_k= z\, h_\la(k)$, $z=z(\t)$ is a function of the background, $h_\la$ are the scalar coefficients in front of the two polarization modes $\la=+,\times$ of the tensor perturbation (in $D=4$ dimensions and in the absence of a mass), and $c_{\rm t}$ is the propagation speed of the perturbation. In Einstein gravity, $z=a$ and $c_{\rm t}^2=1$ \cite{Mukhanov:1990me}. During inflation or for \emph{any} (even noninflationary) power-law expansion, $z=a\propto 1/\t^2$ and the Hubble radius is $H^{-1}=O(1)\,a\tau$, so that $\p_\t^2z/z=\p_\t^2a/a\propto (aH)^2$. Inside the horizon ($\k\t\gg 1$), the solution $v_k(\t)$ is the harmonic oscillator, while outside the horizon ($\k\t\ll 1$), the solution is $v_k\propto z$ and the tensor mode $h_\la$ is constant. Thus, the Hubble horizon marks the transition between an oscillating (quantum) regime inside and a classical one where perturbations are frozen outside.

The same happens in our model because the solutions $E_{\mu\nu}=0$ of Einstein's gravity are also solutions of the equations of motion \Eq{LEOM0} of this nonlocal theory at the classical level and one can apply standard perturbation theory around the radiation-dominated FLRW background. Quantum corrections modify the effective equations of motion as in \Eq{linear0} but do not alter this argument appreciably because they are sub-dominant. Thus, the tensor quantum modes naturally transit to a classical regime in the same way as they do in standard and higher-derivative cosmology, i.e., passing through the horizon. The only difference is that, while in inflation the background accelerates and the transition occurs when perturbations cross the horizon outwards, in the present model the background is decelerating and the transition occurs at horizon re-entry.

This conclusion finds a correspondence in the limit \Eq{modeB}, where the Mukhanov--Sasaki equation of tensor modes is \Eq{musa} with $z=a \sqrt{12\cB(\dot H+H^2)}/\Mpl$ and $c_{\rm t}^2=1-8\cB \dot H/\Mpl^2$ \cite{Noh:1996da,Yun:2020jxx}. Therefore, for a power-law expansion as in our radiation-dominated universe we have again that $\p_\t^2z/z\propto (aH)^2$ and the Hubble horizon (actually, the time-varying sound horizon) still separates an oscillatory and a frozen regime for perturbations ($c_{\rm t}\k\t\gg 1$ and $c_{\rm t}\k\t\ll 1$, respectively).

As a cautionary note, it would be hazardous to take all characteristics of the approximation \Eq{modeB} at face value, even if we found some in apparent agreement with the behaviour of the radiation-domination solution in Einstein gravity, which is also an exact solution of the full nonlocal theory. The problem is that \Eq{modeB} describes an intermediate regime valid only in a certain energy or time range. A careful exploration of the higher-derivative dynamics \Eq{modeB} shows some points of departure with respect to the Einstein-gravity picture, all of which are artefacts from the perspective of the full theory. The expression $c_{\rm t}^2=1-8\cB \dot H/\Mpl^2$ for the propagation speed implies that $c_{\rm t}^2<1$ only when $\cB>0$ \cite{Yun:2020jxx}, which is the opposite case with respect to ours according to \Eq{modeB}. Moreover, for us $|\cB|\propto b$ is very large as per \Eqq{nTconc}. This looks alarming but is actually a spurious effect.  Since the nonlocal form factor in \Eqq{LEOM0} is entire and does not modify the standard dispersion relation, the propagation speed of primordial gravitational waves in the full theory is exactly $c_{\rm t}=1$. For the same reason, we cannot take the solutions of \Eqq{musa} with $z$ and $c_{\rm t}$ given in \cite{Yun:2020jxx} as exact solutions of the theory with the correct stability properties. In the present case, the approximation (which, we repeat, is not a truncation) $F(\Delta)\sim {\rm const}$ introduces a spurious modification of the propagation speed $c_{\rm t}$ but it does not modify qualitatively the idea that cosmological perturbations turn classical when crossing the horizon. Another, perhaps more delicate example of artefact of \Eq{modeB} is that, in this model, the radiation-dominated solution $a=(t/t_0)^{1/2}$ is past-stable against small inhomogeneous and anisotropic perturbations \cite{Barrow:2007pm,Cotsakis:2012yr,Cotsakis:2013aqa,Lehners:2019ibe}, while in Einstein gravity it is not \cite{Misner:1969hg,Belinsky:1970ew,Belinsky:1982pk}. In the first case, the initial conditions are necessarily isotropic, which seems in agreement with the super-selection made by Weyl symmetry in the trans-Planckian phase. This is only a coincidence not to be taken seriously, for two reasons. First, the actual exact FLRW solution of the full classical theory has the same stability properties as in Einstein gravity, where at $t=0$ one meets the Belinsky--Khalatnikov--Lifshitz (BKL) anisotropic singularity \cite{Misner:1969hg,Belinsky:1970ew,Belinsky:1982pk}. Second, one cannot extrapolate the behaviour of the FLRW solution to arbitrarily early times in nonlocal quantum gravity because Weyl invariance removes the big bang and makes FLRW solutions conformally and physically equivalent to Minkowski spacetime, or to solutions interpolating from Minkowski to FLRW \cite[section 5]{Modesto:2022asj}.


\subsection{Scalar perturbations} \label{ScalarPerturbations}

Around \Eqq{raddo}, we made a first mention to the matter content of the Universe as radiation. This choice is dictated by continuity from the Weyl phase, where the dynamics of all matter fields is Weyl-invariant by construction. When this symmetry is spontaneously broken, all fields with nonzero conformal weight acquire couplings or masses through the expectation value of the dilaton. At early times in the Higgs phase, the total energy density of massless particles with zero conformal weight scales as $\sim a^{-4}$ in four dimensions and dominates over dust-like contributions $\sim a^{-3}$ (sectors of fields with nonzero conformal weight). An example of the first is the Maxwell sector, which stays conformally invariant \cite{Cote:2019kbg}. In $D=4$ FLRW spacetime, the homogeneous component of the energy-momentum tensor corresponding to the massless sector is radiation (see, e.g., \cite{Faraoni:2020iel} and references therein), described by a perfect fluid with barotropic index $w=1/3$. Thus, the symmetry structure of the theory at early times spontaneously generates the radiation-dominated era necessary for post-CMB evolution and big-bang nucleosynthesis. The other matter fields of the Standard Model exit their vacua after Weyl symmetry breaking and can later contribute collectively as a dust fluid during the matter-domination era. (We do not address the dark energy problem here.)

Scalar perturbations are thermal fluctuations of the radiation filling the Universe at the beginning of the Higgs phase. They are not induced by the quantum fluctuations of a scalar field as in inflation. In order to convert the power spectrum of primordial density perturbations to the spectrum of fluctuations in the CMB at large angular separations, we follow the standard treatment of the Sachs--Wolfe effect relating temperature deviations $\de T/T$ to the gravitational potential $\Phi$ at the last-scattering surface \cite{Sachs:1967er,White:1997vi}: 
\be
\frac{\de T (\bm{x})}{T}  = -\frac{1+3w}{3(1+w)} \,  \Phi(\bm{x}) \,.
\label{deltaT}
\ee
For radiation, $w=1/3$ and the front coefficient in \Eqq{deltaT} is $-1/2$.

At the last-scattering surface, the theory \Eq{action} is well approximated by Einstein gravity. Therefore, the potential is related to density perturbations by the standard Poisson equation
\be
\N^2 \Phi(\bm{x}) = \frac{1}{2\Mpl^2} \de \rho(\bm{x}) 
\,.
\label{poisson}
\ee
Regarding the potential, at the last-scattering surface the leading contribution to $\Phi$ comes from the Einstein--Hilbert action because, at that time, the energy scale is much smaller than the Planck energy \cite{Antoniadis:1996dj}. Moreover, we remind the reader that the FLRW solutions of Einstein's equations are exact solutions also of the theory \Eq{action}.

Equations \Eq{deltaT} and \Eq{poisson} imply
\ben
\frac{\de T(\bm{x})}{T} = -\frac{1+3w}{3(1+w)} \,  \frac{1}{2\Mpl^2}\frac{1}{\N^2} \de \rho(\bm{x})\,,
\een
so that the two-point function of CMB temperature fluctuations reads
\ba
\hspace{-.9cm}\left\langle \frac{\de T(\bm{x})}{T} \frac{\de T(\bm{y})}{T}\right\rangle &=& \left[\frac{1+3w}{3(1+w)}\right]^2\left\langle \Phi(\bm{x}) \Phi(\bm{y}) \right\rangle,\label{CTheta}\\
\hspace{-.9cm}\left\langle \Phi(\bm{x}) \Phi(\bm{y}) \right\rangle &=& \frac{1}{4\Mpl^4} \left\langle\frac{1}{\N^{2}_{\bf x}}\de\rho(\bm{x}) \frac{1}{\N^{2}_{\bf y}} \de\rho(\bm{y})\right\rangle,\label{ScalarP}
\ea
where we formally inverted \Eqq{poisson}. 

On the other hand, both the density perturbations and their two-point correlation function have to be evaluated at the energy scale $E \simeq \Lst/10$ right after the Weyl phase. Therefore, we need to express the perturbations of the energy-momentum tensor in terms of the gravitational perturbations through the linearized equations of motion for the theory \Eq{action} at the scale \Eq{energyRegime}. Afterwards, such density perturbations affect the fluctuations of the gravitational potential, which satisfies the $00$ component \Eq{poisson} of Einstein's equations at the decoupling time $t_{\rm dec}$. In turn, the potential affects the temperature of the CMB. Schematically, 
\ban
\de\rho\Big|_{E \simeq \frac{\Lst}{10}} & \underset{(\ref{poisson})}{\Longrightarrow}& 
\Phi\Big|_{\mbox{last scattering}}  \Longrightarrow \frac{\de T}{T}\Big|_{\mbox{last scattering}}\,.
\ean
In order to relate the density perturbations to the graviton two-point correlation function, we need the linearized equations of motion for the theory \Eq{action} including the quantum corrections, i.e.,
\be
\left[\rme^{\H^{(0)}}\right]_{\mu\nu}^{\s\t}\left[\frac{\Mpl^2 G^{(1)}_{\s\t}-T^{(1)}_{\s\t}}{2}\right]+E^{Q (1)}_{\mu\nu}=0
\,,
\label{LEOM}
\ee
where $E^Q$ are the quantum corrections to the equations of motion coming from the Lagrangian \Eq{LQ}. The linearized equations are given in appendix \ref{appeB4} by \Eqq{EoMLinT}, written in compact notation as \Eq{EoMQh10}.

Now we are ready to evaluate the two-point correlation function of the energy momentum tensor. The details of this tedious computation are given in appendix \ref{appeE} and the final result is \Eqq{TT}. However, we are only interested in the two-point correlation function of the energy-density perturbation $\de\rho \coloneqq T_{00}^{(1)}$, which is given by \Eqq{roro2}. Let us focus on the first contribution proportional to $5/12$, later showing that the other term is negligible: 
\ba
\langle\de\rho(x)\de\rho(y) \rangle \!\!\!&\simeq&\!\!\! -\frac{4\Mpl^2\Lz^2}{b\,\rme^{\tilde\g_\textsc{e}}}\! \int \! \frac{\rmd^4 k}{(2\pi)^4}\,\rme^{\rmi k\cdot(x-y)}\frac{5}{12}\left(\frac{k}{\Lst}\right)^{2\e_2}\nn
\!\!\!&\stackrel{\textrm{\tiny \Eq{xTOk}}}{=}&\!\!\! -\frac{5\Mpl^2\Lst^{2(1-\e_2)}}{3b\,\rme^{\tilde\g_\textsc{e}}}\,\frac{(|x-y|^2)^{-\e_2-2}}{B_4(2 \e_2 +4)} \,.\label{roro4x}
\ea
On the spatial section $x^0 = y^0$, this becomes
\ba
\langle\de\rho(\bm{x}) \de\rho(\bm{y})\rangle &=& -\frac{5\Mpl^2\Lst^{2(1-\e_2)}}{3b\,\rme^{\tilde\g_\textsc{e}}}\,\frac{(|\bm{x}-\bm{y}|^2)^{-\e_2-2}}{B_4(2 \e_2 +4)}\nn
&=& -\frac{5\Mpl^2\Lst^{2(1-\e_2)}}{3b\,\rme^{\tilde\g_\textsc{e}}}\,\frac{B_3(2 \e_2 +4)}{B_4(2 \e_2 +4)}\int \! \frac{\rmd^3 \bm{k}}{(2\pi)^3}\,\rme^{\rmi \bm{k}\cdot(\bm{x}-\bm{y})}(k^2)^{\e_2+2-\frac{3}{2}}\,,\label{roro4Vecx}
\ea
where in the last equality we used again \Eqq{xTOk} for $D=3$ and $s = \e_2+2$. Using \Eqq{xTOkb}, we can compute the ratio 
\be
\frac{B_3(2 \e_2 +4)}{B_4(2 \e_2 +4)} 
= \frac{\G \left( - \e_2 - \frac{1}{2} \right)}{2 \sqrt{\pi} \G(- \e_2 )} \simeq \e_2\,, 
\ee
where we assumed $\e_2 \ll 1$. Finally, 
\ba
\langle \de\rho (\bm{x}) \de\rho(\bm{y}) \rangle &\simeq&
-\frac{5\Mpl^2\Lst^{2(1-\e_2)}}{3b\,\rme^{\tilde\g_\textsc{e}}}\,\e_2\int \! \frac{\rmd^3 \bm{k}}{(2\pi)^3}\,\rme^{\rmi \bm{k}\cdot(\bm{x}-\bm{y})}k^{2\e_2+1}. 
\label{roro4Vecx2}
\ea
At this point, we can compute \Eqq{ScalarP} employing the definitions of appendix \ref{appeD}. In particular, we apply \Eqq{PGk} to $\psi(\bm{x})=\Phi(\bm{x})$:
\ba
\left\langle\Phi(\bm{x}) \Phi(\bm{y})\right\rangle &=& \frac{1}{4\Mpl^4} 
\left[-\frac{5\Mpl^2\Lst^{2(1-\e_2)}}{3b\,\rme^{\tilde\g_\textsc{e}}}\right]\e_2\int \! \frac{\rmd^3 \bm{k}}{(2\pi)^3}\,\rme^{\rmi \bm{k}\cdot(\bm{x}-\bm{y})} \frac{1}{k^4} k^{2\e_2+1}\nn
&=& -\frac{5\Lst^{2(1-\e_2)}}{12b\,\rme^{\tilde\g_\textsc{e}}\Mpl^2}\frac{\e_2}{2\pi^2} \int_{0}^{+\infty} \frac{\rmd k}{k} \frac{\sin(kr)}{kr} k^{2\e_2} \nn
&=& \int_{0}^{+\infty} \frac{\rmd k}{k} \Delta_\Phi^2(k) \frac{\sin(kr)}{kr} \,,
\ea
where, according to \Eqq{DeltaG}, we introduced the power spectrum
\ba
\Delta_\Phi^2(k) &=& -\frac{5\Lst^2}{6b\,\rme^{\tilde\g_\textsc{e}}\Mpl^2}\frac{\e_2}{(2\pi)^2} \left(\frac{k}{\Lst}\right)^{2\e_2}\nn
&\stackrel{\textrm{\tiny \Eq{CbRic}}}{=}&\frac{5}{12}\frac{\e_2^2}{(2\pi)^2 \b_{\rm Ric}} \left(\lst k\right)^{2\e_2},
\label{Deltas512}
\ea
which is positive since $\b_{\rm Ric}$ is positive. 

In order to get the full result including the second term in \Eqq{roro2} proportional to $5/24$, we simply have to add to \Eqq{Deltas512} a similar contribution obtained replacing $5/12$ with $5/24$ and $\e_2$ with $\e_0$. Hence, the total contribution to the power spectrum is
\ba
\Delta_\Phi^2(k) & = &\frac{5}{12}\frac{\e_2^2}{(2\pi)^2 \b_{\rm Ric}} \left(\lst k\right)^{2\e_2} + \frac{5}{24}\frac{\e_0^2}{(2\pi)^2 (\b_{\rm Ric} + 3\b_R)} \left(\lst k\right)^{2 \e_0}\nn
&=&\frac{5}{12}\frac{\e_2^2}{(2\pi)^2 \b_{\rm Ric}} \left(\lst k\right)^{2\e_2}+\frac{5}{24} \frac{\e_2^2(\b_{\rm Ric} + 3 \b_R)}{(2\pi)^2\b_{\rm Ric}^2} \left(\lst k\right)^{2\e_0}\nn
&=& \frac{5}{12}\frac{\e_2^2}{(2\pi)^2\b_{\rm Ric}} \left[\left(\lst k\right)^{2\e_2}
+\frac{\b_{\rm Ric} + 3 \b_R}{2\b_{\rm Ric}} \left(\lst k\right)^{2\e_0}\right].
\label{CTheta4}
\ea
However, according to \Eqq{bllb}, the coefficient of the second term is $\ll1$. Therefore, \Eqq{Deltas512} is a good approximation of the scalar power spectrum. 

The primordial scalar spectrum $\cP_{\rm s}$ is defined by the spectrum of the gauge-invariant curvature perturbation on uniform density slices $\zeta$, $\cP_{\rm s}(k)\coloneqq k^3P_\zeta(k)/(2\pi^2)$. In the presence of one barotropic fluid with constant $w$ and ignoring the anisotropic stress, one can show that the relation between the potential $\Phi$ and $\zeta$ is (see, e.g., \cite[chapter 8]{LiL})
\be
\Phi\simeq-\frac{3+3w}{5+3w}\,\zeta\,.
\ee
In particular, for radiation $\Phi=-(2/3)\zeta$, so that the final form for the scalar power spectrum is
\be\label{SImplifiedDs}
\boxd{\cP_{\rm s}(k)=\frac{9}{4}\Delta_\Phi^2(k)=\frac{15}{64}\frac{(n_{\rm s}-1)^2}{(2\pi)^2 \b_{\rm Ric}} \left(\lst k\right)^{n_{\rm s}-1},}
\ee
since the scalar spectral index is
\be\label{nS}
\boxd{n_{\rm s}-1\coloneqq\frac{\rmd\ln\cP_{\rm s}}{\rmd\ln k}=2\e_2=-\frac{4\b_{\rm Ric} \Lz^2}{b \,\rme^{ \tilde{\g}_\textsc{e}}\Mpl^2}\,,}
\ee
which is negative definite if $b>0$, as we chose from the start in hindsight. Here we used \Eqq{CbRic}.

Comparing \Eqqs{nT} and \Eq{nS}, we get the prediction
\be\label{enti}
\boxd{n_{\rm t} = 1-n_{\rm s}>0\,.}
\ee
This relation between the tensor and the scalar index is typical of noninflationary scenarios where scalar perturbations are sustained by thermal fluctuations. Examples are string-gas cosmology \cite{BrVa1,Brandenberger:2015kga,Bernardo:2020bpa}\footnote{A noninflationary model coming from matrix theory may realize string-gas cosmology from first principles \cite{Brahma:2021tkh,Brahma:2022hjv}. Its cosmological spectra are under construction.} and the new ekpyrotic model \cite{Brandenberger:2020tcr,Brandenberger:2020eyf,Brandenberger:2020wha}.


\subsection{Parameter space}\label{paraspa}

Let us now discuss the parameter space of the theory compatible with data. 

The theory does not fix the value of the fundamental energy scale $\Lst=\lst^{-1}$ and the only general constraint we can impose is that $\Lst$ be larger than $\sim 10\,{\rm TeV}$ (LHC center-of-mass scale). On the other hand, the bound \Eq{LambdaMp2} gives an upper limit on the energy scale $\Lst$ for the cosmological model in the Higgs phase to be self-consistent, so that overall
\be\label{LambdaMp3}
10\,{\rm TeV} < \Lst < 10 \, \Mpl\,, \qquad \frac{\lp}{10} < \lst < (10\,{\rm TeV})^{-1}\,.
\ee
On the other hand, Weyl invariance leads to the natural limit $\Lst\leq\Mpl$ \cite{Modesto:2022asj}. Instead of using this to shrink the interval \Eq{LambdaMp3}, here we take a conservative stance and consider a worst-case scenario where $\Lst$ is as high as possible, since in quantum gravity one naively expects that the higher the fundamental energy, the more difficult to extract observable effects. Therefore, we consider the narrow interval
\be\label{LambdaMp4}
\Mpl \leq \Lst < 10 \, \Mpl\,, \qquad \frac{\lp}{10} < \lst \leq \lp\,.
\ee
As we will see in section~\ref{tetoscara}, decreasing $\Lst$ (increasing $\lst$) would increase the main effect of the model, which will give us the opportunity to put an observational lower bound on $\Lst$ (upper bound on $\lst$).

Now we discuss the parameter $\b_{\rm Ric}$. According to the \textsc{Planck} Legacy Release \cite{Planck:2018vyg}, at the CMB pivot scale $k_0 = 0.05 \,{\rm Mpc}^{-1}$ and assuming no running, the scalar index is $n_{\rm s}=0.9649 \pm 0.0042$ at 68\% confidence level, assuming $\rmd n_{\rm s}/\rmd\ln k=0$. Plugging the observed value of $n_{\rm s}$ into \Eqqs{DeltaTnT} and \Eq{SImplifiedDs} and taking $\lst=\lp$, we find
\be\label{PtPs}
\hspace{-.2cm}\cP_{\rm t}(k_0) \simeq \frac{8.266 \times 10^{-6}}{\b_{\rm Ric}},\qquad 
\cP_{\rm s}(k_0) \simeq \frac{7.867 \times 10^{-4}}{\b_{\rm Ric}}. 
\ee
Therefore, in order to get the observed amplitude $\cP_{\rm s}(k_0)\approx 2.2 \times 10^{-9}$, from \Eqq{PtPs} it follows that $\b_{\rm Ric} \approx 3.6 \times 10^5$. Such a large value of this coefficient can be justified by noting that it depends on the number of fundamental particles in the matter sector, in which case one might consider to populate the theory with superfields. Since $\b_{\rm Ric} \propto N_{\rm fields}$ (appendix \ref{appeB3} and \cite{BOS,Avramidi:2000bm}), where $N_{\rm fields}$ is the total number of fields in the theory, the latter requires a large number of new fundamental particles. This assumption may be consistent with a particle-physics solution of the problem of dark matter, interpreted as the contribution of exotic particles, while at the same time it could also populate the ``desert'' lying between the mass of the Higgs boson and the Planck scale.

The theory does not predict the value of the free mass scale $\Mpl/\sqrt{b}$ that appears in $n_{\rm t}$, but we can argue that it takes an intriguing value. From \Eqqs{nT} and \Eq{enti} with $\Lz=\Mpl$,
\be
b =\frac{4\b_{\rm Ric}}{(1-n_{\rm s})\,\rme^{\tilde{\g}_\textsc{e}}}\,.
\label{nTconc}
\ee
For $\b_{\rm Ric} \approx 3.6 \times 10^{5}$ and $\tilde{\g}_\textsc{e}=\g_\textsc{e}$, we get $b \approx 2.3 \times 10^7$ and an effective intermediate mass scale $\Mpl/\sqrt{b}=5 \times 10^{14}\,{\rm GeV}$, which looks like a low-end grand-unification scale and is about one order of magnitude larger than the Starobinsky mass for the curvaton. Grand unification scenarios are much easier to realize in the presence of supersymmetry, which would naturally generate a large $\b_{\rm Ric}$.

The scenario just described is not the only possibility to explain the observed value of the scalar spectral amplitude. Remaining within the non-supersymmetric Standard Model of particle physics, another mechanism could be rooted in the choice of a different background in the post-Planckian phase, in particular, the interpolating metric \cite{Modesto:2022asj}
\be
g_{\mu\nu}' = - \left( A + B \tanh \frac{\t}{\t_\Pl}\right)\eta_{\mu\nu}, \qquad
\lim_{\t \rightarrow \pm \infty} a^2 = A \pm B\,, \qquad A \geq B\,,\label{BD}
\ee
where $\t$ is conformal time. When performing a Weyl rescaling $\eta_{\mu\nu}=\Om^{-2} g_{\mu\nu}'$ with $\Om=a(\tau)$ from the Minkowski metric $\eta_{\mu\nu}$ to the spacetime \Eq{BD}, the two-point correlation function must be rescaled, too. Since the scalar potential $\Phi$ is proportional to the metric perturbation $\bm{h}$ and $\bm{h}^\prime = \Om^2 \bm{h}$, one has $\langle\Phi(\bm{x}) \Phi(\bm{y})\rangle' = \Om^4 \langle\Phi(\bm{x}) \Phi(\bm{y})\rangle$ and
\be
\cP_{\rm s}^\prime = \Om^4 \cP_{\rm s}\,. 
\ee
The same rescaling is applied to the amplitude $\cP_{\rm t}$, but the ratio $r\coloneqq \cP_{\rm t}/\cP_{\rm s}$ is Weyl-invariant. Therefore, evaluating $\Om^2$ at the Planck conformal time, i.e., for $\t \gtrsim \t_\Pl$, $\Om^2(\t_\Pl) = a^2(\t_\Pl) \simeq A+B$, we get an amplitude $\cP_{\rm s}^\prime(k_0)$ consistent with observations if, for instance, $\b_{\rm Ric}=1$ and $A+B = 1.7\times 10^{-3}$, or $\b_{\rm Ric}=40$ and $A+B=10^{-2}$. Therefore, in this case the level of fine tuning is very modest.


\subsection{Tensor-to-scalar ratio}\label{tetoscara}

Using \Eqqs{DeltaTnT} and \Eq{SImplifiedDs}, we can compute the tensor-to-scalar ratio at the pivot scale $k_0$:
\be\label{tsr}
\boxd{r \coloneqq \frac{\cP_{\rm t}(k_0)}{\cP_{\rm s}(k_0)} =\frac{64}{15(1-n_{\rm s})}\left(\lst k_0\right)^{2(1-n_{\rm s})}.}
\ee
Remarkably, this observable only depends on the scale $\lst$, while the other parameters of the model, the constants $\b_{\rm Ric}$ and $b$, are hidden in the scalar index. This implies that, despite the need of more understanding about the values of these constants, the theory is already falsifiable. In fact, from \Eqq{LambdaMp2} and for $0.1\leq \lst/\lp \leq 1$, we obtain the very narrow range
\be\label{r005}
0.009\leq r \leq 0.011\,, \qquad k_0 = 0.05 \,{\rm Mpc}^{-1}\,,
\ee
with the value
\be\label{r005b}
\boxd{r_{0.05}=0.011\qquad (\Lst=\Mpl)}
\ee
being the most natural one ($\lst=\lp$) and
\be\label{r005c}
\boxd{r_{0.05}\geq 0.009}
\ee
being the \emph{lower bound} on $r$ predicted by the theory. Not only are these numbers three times larger than the prediction of Starobinsky inflation \cite{Starobinsky:1980te,Akrami:2018odb} ($r_{\rm Starobisky} = 0.0037$ for $\cN=57$ e-folds) or Weyl-squared quantum gravity \cite{Anselmi:2020lpp}, but they are also within reach of present observations on primordial gravitational waves \cite{BICEP:2021xfz,Tristram:2021tvh}. BICEP Array, the imminent upgrade of Keck Array and BICEP3, is projected to reach a sensitivity allowing to determine the tensor-to-scalar ratio with an uncertainty $\s(r)\sim 0.003$ by 2027 \cite{BICEP:2021xfz,BICEPKeck:2024stm}. Therefore, this instrument will be able to detect any signal with $r>0.009$, establish an implication for $r\sim 0.006\mhyp 0.009$, and rule out any model predicting $r>0.006$ in the case of no signal being found. The satellite LiteBIRD will further reduce the uncertainty in case of detection \cite{LiteBIRD:2022cnt,LiteBIRD:2023iei}. In contrast, Starobinsky inflation will have to wait LiteBIRD to be verified, which will establish an upper bound of $r<0.002$ at the 95\% confidence level in case of no detection.

Since the larger $\lst$ the larger $r$, we can extract an upper bound on $\lst$ (lower bound on $\Lst$) from \Eqq{tsr} and the \textsc{Planck}-BICEP-Keck observational constraint $r_{0.05}<0.036$ at 95\% confidence level \cite{BICEP:2021xfz}:
\be\label{newbound}
\Lst> 8.5\times 10^{10}\,{\rm GeV}\,,
\qquad \lst< 2.3\times 10^{-27}\,{\rm m}\,.
\ee
This lower bound on $\Lst$ is a dramatic improvement with respect to the $O({\rm TeV})$ bound from LHC \cite{Biswas:2014yia}. It also highlights yet another advantage of this model with respect to inflation. The latter is affected by a theoretical puzzle consisting in the mismatch between the initial energy scale of the universe $H\sim \Mpl$ expected from quantum physics and the energy scale of inflation at horizon crossing $H_{\rm infl}< 10^{-5}\Mpl$ as determined by observations \cite{Akrami:2018odb}. There exist ways to reconcile this discrepancy, such as delaying the onset of inflation or making a suitable choice of vacuum (see \cite{Kowalczyk:2024ech} and references therein), but in our case there actually is no theoretical tension. The above upper bound on the inflationary scale $H_{\rm infl}$ comes from the upper bound on $r$, which for us results in the \emph{lower} bound \Eq{newbound} on the fundamental scale $\Lst$. This is compatible with the initial energy $H\sim \Mpl$ dictated by the scale of Weyl symmetry breaking \Eq{Wesib}. A natural configuration would be $H\sim\Lst=\Mpl$.


\subsection{Primordial GW background}

\begin{figure}[ht]
	\centering
\includegraphics[width=15cm]{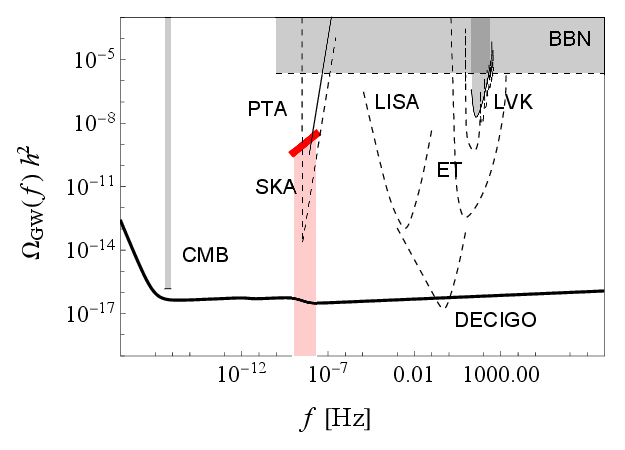}
\caption{Gravitational-wave background \Eq{Omgw} with primordial tensor spectrum \Eq{DeltaTnT} compared with the sensitivity curves (dashed) of LIGO-Virgo-KAGRA (LVK; O2 and designed sensitivity curves), SKA, LISA, Einstein Telescope (ET; 10 km triangular configuration with a signal-to-noise ratio 1 and a one-year observation run \cite{Branchesi:2023mws,ETsensitivity}), and DECIGO. Here $\lst=\lp$; other choices close to the Planck scale do not change the curve appreciably. The single-frequency CMB constraint has been given thickness to make it more visible in the plot. Also shown are the big-bang nucleosynthesis (BBN) exclusion region and, as a thick red segment, a model of the signal (compatible with a population of supermassive black-hole binaries) recently detected by the international network of Pulsar Timing Arrays \cite{NANOGrav:2023gor,EPTA:2023fyk,Reardon:2023gzh,Xu:2023wog,InternationalPulsarTimingArray:2023mzf}. Stochastic backgrounds weaker than the observed signal would be drowned by it at those frequencies (shaded red region). However, the high-frequency cutoff of this signal is much smaller than the DECIGO range.\label{fig2}}
\end{figure}

Since
\be
\e_2 = \frac{n_{\rm s}-1}{2} \approx - 0.01755 \,, \qquad k_0 = 0.05 \,{\rm Mpc}^{-1}\,,
\ee
the primordial tensor spectrum is blue-tilted with a spectral index
\be
n_{\rm t} \approx 0.0351\,.
\ee
This can give rise to an observable gravitational-wave background. The theory reduces to Einstein gravity soon after the Weyl symmetry breaking and by the time cosmological perturbations re-enter the horizon. Therefore, one can use the standard dimensionless spectral shape
\be\label{Omgw}
\hspace{-.2cm}\Om_\textsc{gw}(k,\t_0)\coloneqq\frac{1}{\rho_{\rm crit}} \frac{\rmd\rho_\textsc{gw}}{\rmd \ln k} =\frac{k^2}{12a_0^2H_0^2}\cP_{\rm t}(k)\,\cT^2(k, \tau_0),
\ee
where $\rho_{\rm crit}=3\Mpl^2H_0^2$ is the critical energy density, $\rho_\textsc{gw}$ is the energy density of gravitational waves (spatial average of the kinetic energy $\dot h^2$ of the perturbation), the subscript $0$ denotes quantities evaluated today, $\tau$ is conformal time, and $\cT$ is a transfer function describing how the primordial spectrum evolved since its generation \cite{Turner:1993vb,Watanabe:2006qe,Kuroyanagi:2008ye,Nakayama:2008wy,Kuroyanagi:2014nba}. Its form, valid for any theory where the observed perturbations came from outside the horizon (independently of where they were generate, in- or outside) can be found, e.g., in \cite{Calcagni:2020tvw}. This formula applies also to our non-inflationary scenario where the universe has always been radiation dominated and there is no reheating, a situation which can be assimilated to instantaneous reheating at the Planck temperature $T_{\rm reh}=T_\Pl$. Then, as shown in figure~\ref{fig2}, the stochastic background will be able to reach the sensitivity curve of the planned DECIGO interferometer \cite{Kawamura:2020pcg}, thus providing an independent test of the theory.


\section{Conclusions}\label{conclu}

In this paper, we have continued the analysis of the cosmological scenario of the early universe arising in nonminimally coupled asymptotically local quantum gravity started in \cite{Modesto:2022asj}. In this model, generated directly in the full theory under a minimal set of approximations, we have calculated the tensor and scalar primordial spectra, their indices, the tensor-to-scalar ratio, and the gravitational-wave background. These findings rely on the property of finiteness of the theory, that secures Weyl invariance at the classical as well as at the quantum level. Once Weyl symmetry is spontaneously broken, two-point correlation functions get perturbative logarithmic quantum corrections that make primordial spectra deviate from exact scale invariance. 

The main theoretical results of the paper are the tensor spectrum $\cP_{\rm t}$ \Eq{DeltaTnT}, the tensor index $n_{\rm t}$ \Eq{nT}, the scalar spectrum $\cP_{\rm s}$ \Eq{SImplifiedDs}, the scalar index $n_{\rm s}$ \Eq{nS}, the relation \Eq{enti} between the tensor and the scalar index, and the tensor-to-scalar ratio $r$ \Eq{tsr}. The tensor spectral index $n_{\rm t}$ is positive and equal to $1-n_{\rm s}$ and the tensor-to-scalar ratio $r$ only depends on $n_{\rm s}$ and the fundamental length scale $\lst$.

The main phenomenological results are the lower bound \Eq{r005c} for $r$ (and its most natural value \Eq{r005b}), and the primordial gravitational-wave background shown in figure~\ref{fig2}. These are unique, falsifiable predictions of the theory. Moreover, we have obtained the strong lower bound \Eq{newbound} on the energy scale $\Lst$, seven orders of magnitude higher than previous bounds around the LHC center-of-mass energy.

The only free parameters in the model are the constants $\b_{\rm Ric}$ (or, actually, the number of matter fields in the spectrum of the theory) and $b$ (or, equivalently, $n_{\rm s}$, which is known from observations). However, $\b_{\rm Ric}$ can only affect the amplitude of the perturbations, but not $n_{\rm t}$ or $r$, that are completely fixed in our model regardless of the details of the theory in the UV regime. The whole scenario is summarized in figure~\ref{fig1}.

If observations failed to detect a large tensor-to-scalar ratio at the level predicted by our model, the latter should be ruled out because there are no really adjustable parameters in \Eqq{tsr}, since the scale $\lst$ cannot vary  more than one order of magnitude within the allowed range. Indeed, inequalities \Eq{LambdaMp2} are not imposed for convenience but are dictated by theoretical arguments during the Weyl phase and by the self-consistency of the perturbative treatment of the quantum field theory. As a consequence, the value of $r$ cannot change by more than 10\%, as seen in \Eq{r005}.

One cannot give up any of the assumptions and approximations without destroying the whole set of predictions. These approximations are only three: quadratic Stelle-like limit, validity of the perturbative expansion, and flat spacetime regime. Any attempt to relax our construction to something else adapting to future observations (for instance, a lower or undetected $r$) would inevitably lead to a more complicated model with more assumptions, or to the necessity to include an inflaton. This rigidity is due to using perturbation theory in a well-defined way, which has undergone several cross-checks in our treatment. We do not see how one could possibly relax \Eq{LambdaMp2} and similar constraints while keeping the rest of the construction standing. Giving up the higher-derivative regime would not only make the problem possibly intractable, but it would also break the quasi-scale invariance of the primordial spectra, for which quadratic curvature operators are necessary in the absence of inflatonic fields. Modifying the model for the sake of saving it would possibly be theoretically unacceptable and, even if one could manage to do so, it would jeopardize the status of the full theory. The latter would not be ruled out \emph{per se}, but it would be removed as a parsimonious contender among other top-down approaches having a say on the early universe.
	
The blue-tilted GW background, not achievable in single-field inflation, can help to discriminate our model from others with a high-$r$ prediction, such as $\a$-attractors and $V\propto\phi^{4/3}$ inflation \cite{Akrami:2018odb}. In parallel, none of the other quantum-gravity-related models with a similar prediction of the GW background has the same level of $r$ \cite{Calcagni:2020tvw}. Therefore, having two independent observations of the GW background at two different frequency ranges is already sufficient to discriminate among models predicting a similar amplitude in either frequency range. 

In order to complete the cosmological picture and further test the predictions of our model, one should look at other aspects of the evolution of the early as well as of the late universe. In the first case, at an observational level, scenarios based on Weyl invariance are known to be able to generate an observable level of primordial non-Gaussianity  \cite{Antoniadis:1996dj,Antoniadis:2011ib,Agrawal:2020xek}. In particular, in \cite{Antoniadis:2011ib,Agrawal:2020xek} it was found that the level of primordial local non-Gaussianity is $|f_{\rm NL}^{\rm local}|=O(1)$, i.e., larger than in standard inflation (where $|f_{\rm NL}^{\rm local}|=O(\e)$ and $\e$ is the first slow-roll parameter) but small enough to be still compatible with the current $1\s$ constraint $f_{\rm NL}^{\rm local}=-0.9\pm 5.1$ \cite{Planck:2019kim}. Since Weyl invariance washes away many of the fine details of the underlying model, it is quite possible that also our theory will produce a level of non-Gaussianity of such a magnitude. None of the models in \cite{Antoniadis:2011ib,Agrawal:2020xek} predict a high $r$, so that non-Gaussianity can provide a genuine cross-check of our model against high-$r$ inflationary scenarios. The calculation of the bispectrum is conceptually straightforward but can be quite involved, and we leave it for future work. This could also shed some light on the decoherence mechanism mentioned in section~\ref{TensorPerturbations}.

At the level of late-time cosmology, in this paper we have said nothing about the cosmological constant problem, which could be another field of application of the symmetries of the theory. Also, any late-time modification of the $\Lambda$CDM model should be able to explain the $H_0$ and the $\s_8$ tensions and satisfy certain general conditions outlined in \cite{Heisenberg:2022gqk,Heisenberg:2022lob,Lee:2022cyh}.

The theory of quantum gravity proposed here directly generates a cosmological model with very few free parameters and rigid, clearly falsifiable predictions in the tensor sector. These will be tested by data on primordial gravitational waves, both in the present (by 2027, BICEP Array) and in the future ($\lesssim 10$ years, LiteBIRD; $\gtrsim 20$ years, DECIGO). Already before next-generation observatories become operational, gravitational-wave astronomy can arise as one of the most powerful tools to probe gravity at the interface with the quantum.


\section*{Acknowledgments}

The authors are supported by grant PID2020-118159GB-C41 funded by the Spanish Ministry of Science and Innovation MCIN/AEI/10.13039/501100011033. L.M.\ was also supported by the Basic Research Program of the Science, Technology, and Innovation Commission of Shenzhen Municipality (grant no.\ JCYJ20180302174206969).


\appendix


\section{Tree-level graviton propagator}\label{appeB}

In this appendix, we calculate the tree-level graviton propagator for the theory \Eq{action} in vacuum, after recalling some results in the literature of quantum gravity. 






\subsection{Tree-level propagator in the theory \Eq{action}}\label{appeB2}

The graviton propagator for the theory \Eq{action} has been computed in \cite{Modesto:2021ief}. Here, we recall the form of the kinetic operator and the gauge-invariant components of the propagator. In $D$ dimensions, the expansion of the vacuum action at second order in the graviton perturbation is
\ba
\hspace{-1.cm}S^{(2)}[h] &=&
\frac{1}{2} \int   \rmd^D x \left\{h^{\a \b}  \, \left( \frac{\de^2 S_{\rm loc}  }{\de h^{\a \b} \de h^{\g \s}} \right) 
\left[\rme^{\bar{\H}(\Hes)}\right]^{\g\s}_{\mu\nu}\,h^{\mu \nu }\right\}\nn
&=& \frac{\Mpl^{2}}{8} \int   \rmd^D x \, 
h^{\a \b}  \,  \left\{\left[ P^{(2)} - (D-2) P^{(0)} \right] \B 
\right\}_{\a\b}^{\g \s} \left\{\rme^{\bar{\H}\left[\frac{P^{(2)} - (D-2) P^{(0)}  }{\Lst^2} \B  \right]} \right\}_{\g \s,\mu \nu}  \, 
h^{\mu\nu} \label{expH} \\
&=& \frac{\Mpl^{2}}{8} \int\rmd^D x \, 
h^{\a\b}\,\left(\left\{P^{(2)}\rme^{\bar{\H}\left(\frac{\B}{\Lst^2} \right)}
- (D-2) P^{(0)}  \rme^{\bar{\H} \left[-\frac{(D-2)\B}{\Lst^2}\right]}\right\}\B\right)_{\a\b,\mu\nu}\, h^{\mu\nu}\nn
& \eqqcolon & \frac{\Mpl^{2}}{8} \int   \rmd^D x \, 
h^{\a\b}  \,  \left\{\left[P^{(2)}  \,\rme^{\H_2}-(D-2)P^{(0)} \,\rme^{\H_0}
\right]\B\right\}_{\a\b,\mu\nu}\,h^{\mu\nu}\nn
& = & \frac{1}{2} \int\rmd^D x\,h^{\a \b}  \,  \cO^{\rm K}_{\a\b\mu\nu}\, h^{\mu \nu}\,,\label{Actiongravity2}
\ea
where $\bar{\H}(z) = \H(z) - \H(0)$ was defined in \Eqq{FF}, the entire function $\H(z)$ is defined in \Eqq{H}, in the second and third equalities we wrote explicitly the dimensionless arguments of $\H$, and we have used the projectors
\bs\label{projectors}\ba
&& P^{(2)}_{\mu\nu\rho\s} \coloneqq \frac{1}{2} \left(\Theta_{\mu \rho} \Theta_{\nu \s} +  \Theta_{\mu\s} \Theta_{\nu\rho} \right)
- \frac{1}{D-1} \Theta_{\mu\nu} \Theta_{\rho\s} \,,\\
&&  P^{(0)}_{\mu\nu\rho\s} \coloneqq \frac{1}{D-1} \Theta_{\mu \nu} \, \Theta_{\rho \s}\,,\qquad \Theta_{\mu \nu} \coloneqq \eta_{\mu \nu} - \om_{\mu\nu}\,.
\ea\es

Therefore, the propagator can be obtained from the propagator of the Einstein--Hilbert theory under the replacements
\ba
&& \B \,  P^{ (2) }\quad \rightarrow \quad \B\,\rme^{\bar{\H}\left(\frac{\B}{\Lst^2}\right)} \,P^{(2)}\eqqcolon \B \, \rme^{\H_2} \, P^{(2)}\,,\label{P2id}\\
&& \B \, P^{ (0) }\quad \rightarrow \quad \B \, \rme^{\bar{\H}\left[-\frac{(D-2) \B}{\Lst^2}\right]}\,P^{(0)} \eqqcolon \B\,\rme^{\H_0}\, P^{(0)}\,.\label{P0id}
\ea
These rules are useful when computing the propagator for the action \Eq{action} including the logarithmic one-loop quantum corrections (see \Eqq{h2Q}). Augmenting the kinetic operator $\mathcal{O}^{\rm K}$ with a gauge-fixing term $\cO_{\rm gf}$, from $\cO = \cO^{\rm K} + \cO_{\rm gf}$ we get the tree-level graviton propagator up to gauge-dependent contributions \cite{Calcagni:2024xku,Accioly:2002tz}:  
\be
\cO^{-1} = \frac{4}{\Mpl^{2}}\left\{\frac{P^{(2)}}{\B \, \rme^{\bar{\H} \left(  \frac{\B}{\Lst^2}\right)}} - \frac{P^{(0)}}{\B \,(D-2)  
	\rme^{\bar{\H}\left[-\frac{(D-2)\B}{\Lst^2}\right]}}\right\}
=\frac{4}{\Mpl^{2}}\left[\frac{P^{(2)}}{\B \,\rme^{\H_2}} - \frac{P^{(0)}}{(D-2)\B\,\rme^{\H_0}}\right]\,.\label{PropAction}
\ee
According to \Eqq{UVlimit}, in the UV the propagator becomes polynomial:
\ba
\cO^{-1} \stackrel{\textsc{uv}}{\simeq} \frac{4}{\Mpl^{2}\rme^{\tilde{\g}_\textsc{e}}}\left\{\frac{P^{(2)}}{\B \, p\left(\frac{\B}{\Lst^2}\right)} - \frac{P^{(0)}}{(D-2)\B\,p \left[-\frac{(D-2)\B}{\Lst^2}\right]}\right\}, \label{PropPoly}
\ea
where $\tilde{\g}_\textsc{e}$ is defined in \Eqq{CbRic}. 
 Writing explicitly the scale in the argument of the polynomial \Eq{Poly},
\ba
&& p\left(\frac{\B}{\Lst^2}\right) = b_0 - b \, \frac{\B}{\Lz^2}+\left(\frac{\B}{\Lst^2}\right)^4\,,\label{poly2}\\
&& p\left[-\frac{(D-2)\B}{\Lst^2}\right] = b_0 + (D-2) b \frac{\B}{\Lz^2}+  \left[(D-2)\frac{\B}{\Lst^2}\right]^4\,.\label{poly0}
\ea
When $b_0\neq 0$, this constant can be dropped from the polynomial at the energy scale $E \gtrsim \Est$ defined in \Eqq{Estar}.


\subsection{One-loop graviton propagator} \label{appeB3}

In this section, we calculate the graviton propagator for the quantum effective Lagrangian \Eq{modeB}, at UV energies where the dynamics of the theory is still dominated by the nonlocal form factors but in such a way that the Lagrangian features a finite number of derivatives. Taking the sum of \Eqq{Actiongravity2} and the quadratic expansion in the graviton $h_{\mu\nu}$ of \Eqq{LQ}, the quantum effective action reads
\ba
\hspace{-1cm}\G^{(2)}[h] &=& \frac{1}{2} \int \rmd^D x \, h_{\mu\nu} \left[ \mathcal{O}^{\rm K} + \mathcal{O}^Q \right]^{\mu\nu \rho\s} h_{\rho \s}\nn
\hspace{-1cm}&=& \frac{\Mpl^{2}}{8}\int\rmd^D x \,  h_{\mu\nu}  \, \B \left( \left(\rme^{\H_2} + \B \g_2^{Q}\right)\, P^{(2)} \right.\nn
&&\qquad\left.-\left\{(D-2)\rme^{\H_0}-\B\left[4(D-1) \g_0^{Q}+ D \g_2^{Q}\right] \right\}P^{(0)}\right)^{\mu\nu \rho\s}h_{\rho \s}, 
\label{h2Q}
\ea
where $\g_2^Q$ and $\g_0^Q$ are the quantum corrections to the kinetic operator. According to \Eqq{LQ}, the quantum form factors are
\be
\g_2^{Q}  \coloneqq \frac{2\b_{\rm Ric}}{\Mpl^2}\,\ln\left(-\frac{\B}{\Lst^2}\right),\qquad \g_0^{Q} \coloneqq \frac{2\b_R}{\Mpl^2}\,\ln\left(-\frac{\B}{\Lst^2 \, \de_0^2}\right).\label{gammaQAB}
\ee
The linearized equations of motion that we get from \Eqq{h2Q} are
\ba
0&=& \B \left(\left(\rme^{\H_2} + \B \,\g_2^{Q}\right)\,P^{(2)} 
-\left\{(D-2)\rme^{\H_0}-\B\left[4(D-1) \g_0^{Q}+ D \g_2^{Q}\right]\right\}P^{(0)}\right)_{\mu\nu}^{\rho \s} h_{\rho \s}\nn
&=& \B \left\{\rme^{\H_2} P^{(2)} - (D-2)\rme^{\H_0} P^{(0)} + \B \g_2^{Q}\, P^{(2)} + \B \left[4 (D-1) \g_0^{Q}+ D \g_2^{Q}\right] 
P^{(0)}\right\}_{\mu\nu}^{\rho \s} h_{\rho \s}\nn
&=& \B \left\{\rme^{\bar{\H}} \left[ P^{(2)} -(D - 2) P^{(0)} \right]+ \B \g_2^{Q}\, P^{(2)} + \B \left[4(D-1)\g_0^{Q} + D \g_2^{Q}\right] 
P^{(0)}\right\}_{\mu\nu}^{\rho \s} \, h_{\rho \s}\nn
&=& \B \, \left( \rme^{ \bar{\H}} \left\{\left[P^{(2)} - (D - 2)  P^{(0)} \right]+ \rme^{-\H_2} \B \g_2^{Q}\, P^{(2)}\right.\right.\nn
&&\qquad\left.\left. 
+ \rme^{-\H_0} \B \left[4 (D-1) \g_0^{Q} + D \g_2^{Q}\right]P^{(0)}\right\}\right)_{\mu\nu}^{\rho \s} \, h_{\rho \s}\,,\label{EoMQh}
\ea
where the operator $\bar\H$ acting on the first term contains the projectors according to \Eqq{expH}. In the last two steps of \Eqq{EoMQh}, we used the completeness relations of the projectors \cite{Riv64,Bar65,VanNieuwenhuizen:1973fi}:
\be\label{copro}
\bm{P}^{(2)^2} = \bm{P}^{(2)}\,,\qquad \bm{P}^{(0)2}= \bm{P}^{(0)}\,,\qquad \bm{P}^{(2)}\bm{P}^{(0)} = 0\,.
\ee 
From the action \Eq{h2Q} augmented by the gauge-fixing term, 
the one-loop one-particle-irreducible propagator for the quantum effective Lagrangian \Eq{modeB} reads
\ba
\bm{\cO}^{-1}(k) &=& -\frac{4}{\Mpl^{2}}\left(\frac{\bm{P}^{(2)}}{k^2 \left[\rme^{\H_2} - k^2 \g_2^{Q} (k^2) \right]  } \right.\nn
&&\qquad\qquad \left.- \frac{\bm{P}^{(0)}}{k^2 \left\{(D- 2)\rme^{\H_0} + k^2 \left[4 (D-1) \g_0^{Q} (k^2)  + D \g_2^{Q} (k^2)\right] \right\}}\right). 
\label{PQEA} 
\ea

In the high-energy limit, we can replace the form factors with the polynomial as we have done in \Eqq{PropPoly}, and, thus, the propagator \Eq{PQEA} turns into
\ba
\bm{\cO}^{-1}\!\! &\stackrel{\textsc{uv}}{\simeq}&\!\! -\frac{4}{\Mpl^{2}}\left(\frac{\bm{P}^{(2)}}{k^2 \left[\rme^{\tilde{\g}_\textsc{e}} p\left(-\frac{k^2}{\Lst^2}  \right) - k^2 \g_2^{Q} (k^2)\right]}\right.\nn
&&\qquad\qquad \left. - \frac{\bm{P}^{(0)}}{k^2 \left\{(D-2)\rme^{\tilde{\g}_\textsc{e}}p\left[\frac{(D-2)k^2}{\Lst^2}\right]+k^2 \left[4 (D-1) \g_0^{Q} (k^2)  + D \g_2^{Q} (k^2)  \right] \right\}}\right)\!. \nn
&& 
\label{PQEA2} 
\ea
Assuming now to be in the energy regime \Eq{interm}, which is still well within what we consider the UV regime, the action reduces to Stelle's theory with a certain choice of coefficients \cite{Modesto:2022asj} 
 and the dominant contribution comes from the first monomial in \Eqqs{poly2} and \Eq{poly0}, since at energy scales $E \gtrsim \Est$ we can forget the constant $b_0$ in the polynomial:
\be
\rme^{\tilde{\g}_\textsc{e}} p\left(-\frac{k^2}{\Lst^2}\right) \simeq \frac{\rme^{\tilde{\g}_\textsc{e}} b \,k^2}{\Lz^2}\,,\qquad \rme^{ \tilde{\g}_\textsc{e}} p\left[\frac{(D-2)k^2}{\Lst^2}\right] \simeq -\frac{\rme^{\tilde{\g}_\textsc{e}} (D-2) \, b \, k^2}{\Lz^2}
\,. 
\ee
The propagator turns into: 
\ban
\bm{\cO}^{-1}(k) & \simeq & -\frac{4}{\Mpl^{2}}\left(\frac{\bm{P}^{(2)}}{k^4 \left[\frac{\rme^{\tilde{\g}_\textsc{e}} b}{\Lz^2} -\g_2^{Q} (k^2)\right]}\right.\nn
&&\qquad\qquad \left. - \frac{\bm{P}^{(0)}}{k^4 \left\{-(D-2)\frac{\rme^{\tilde{\g}_\textsc{e}} (D-2) b}{\Lz^2}+\left[4 (D-1) \g_0^{Q} (k^2)  + D \g_2^{Q} (k^2)  \right] \right\}}\right) \\
&=& -\frac{4}{\rme^{\tilde{\g}_\textsc{e}} b}\frac{\Lz^2}{\Mpl^{2}}\left(\frac{\bm{P}^{(2)}}{k^4 \left[1 -\frac{\Lz^2}{\rme^{\tilde{\g}_\textsc{e}} b}\g_2^{Q} (k^2)\right]}\right.\nn
&&\qquad\qquad\qquad \left. + \frac{\bm{P}^{(0)}}{k^4 \left\{(D-2)^2-\frac{\Lz^2}{\rme^{\tilde{\g}_\textsc{e}} b}\left[4 (D-1) \g_0^{Q} (k^2)  + D \g_2^{Q} (k^2)  \right] \right\}}\right).
\ean
Replacing the quantum corrections \Eq{gammaQAB} in the above expression, we get
\ba
\bm{\cO}^{-1}(k)\!  &=&\! -\frac{4}{\rme^{\tilde{\g}_\textsc{e}} b}\frac{\Lz^2}{\Mpl^{2}} \left(\frac{\bm{P}^{(2)}}{k^4\left[1 - \frac{2\Lz^2}{\rme^{\tilde{\g}_\textsc{e}} b} \frac{\b_{\rm Ric}}{\Mpl^2} \, \ln \left(\frac{k^2}{\Lst^2}\right) \right]}\right. \nn
\!&&\! \left.\hspace{2cm} + \frac{\bm{P}^{(0)}}{k^4 \left\{(D-2)^2 -\frac{\Lz^2}{\rme^{\tilde{\g}_\textsc{e}} b}
	\left[8 (D-1) \frac{\b_R}{\Mpl^2} \, \ln \left(\frac{k^2}{\Lst^2\, \de_0^2}\right) + 2D \frac{\b_{\rm Ric}}{\Mpl^2} \,\ln \left(\frac{k^2}{\Lst^2}\right) \right] \right\}}\right) \nn
\!&=&\! - \frac{4}{\rme^{\tilde{\g}_\textsc{e}} b}\frac{\Lz^2}{\Mpl^{2} k^4}
\left\{ \frac{\bm{P}^{(2)}}{1+ \e_2 \ln \left(\frac{k^2}{\Lst^2} \right)}
+ \frac{\bm{P}^{(0)}}{(D-2)^2  \left[1+ \e_0\ln \left(\frac{k^2}{\Lst^2}\right)
	+\frac{8(D-1) \,  \Lz^2 \, \b_R}{(D-2)^2 \rme^{\tilde{\g}_\textsc{e}} b \, \Mpl^2} \,   
	\ln \de_0^2\right] }\right\}\nn
\!&=&\! - \frac{4}{\rme^{\tilde{\g}_\textsc{e}} b}\frac{\Lz^2}{\Mpl^{2}k^4}
\left[\frac{\bm{P}^{(2)}}{1+ Q_2} + \frac{\bm{P}^{(0)}}{ (D- 2)^2  \left(1+ Q_0  + Q_b \right)}\right],\label{PQEA3A2}
\ea
where we defined
\ba
&& Q_2 \coloneqq \e_2 \ln \frac{k^2}{\Lst^2} \, , \qquad 
Q_0 \coloneqq \e_0 \ln \frac{k^2}{\Lst^2}  \, , \qquad 
Q_b \coloneqq \frac{8 (D-1) \,  \Lz^2 \, \b_R}{(D-2)^2 \, \rme^{\tilde{\g}_\textsc{e}} b \, \Mpl^2} \, \ln \de_0^2 \, , \label{Q2Q0} \\
&&  \e_2 \coloneqq - \frac{2\b_{\rm Ric} \Lz^2}{b \, \rme^{\tilde{\g}_\textsc{e}} \Mpl^2 } \, , \qquad   \e_0 \coloneqq - \frac{2[D \b_{\rm Ric} + 4 (D-1) \b_{ \rm R}] \Lz^2}{(D-2)^2 \, b \, \rme^{\tilde{\g}_\textsc{e}} \Mpl^2} \, .
\label{epsilons}
\ea
In $D=4$ dimensions, \Eqq{epsilons} reads
\ba
\hspace{-.8cm}\e_2 = -\frac{2\b_{\rm Ric} \Lz^2}{b \,\rme^{\tilde{\g}_\textsc{e}}\Mpl^2} \, , \!\qquad   \e_0 = - \frac{2(\b_{\rm Ric} +3 \b_{ \rm R}) \Lz^2}{b \,\rme^{\tilde{\g}_\textsc{e}}\Mpl^2} \, , \!\qquad \e_0 - \e_2 = - \frac{6\Lz^2}{b\,\rme^{\tilde{\g}_\textsc{e}}\Mpl^2} \b_R
= 3 \e_2 \frac{\b_R}{\b_{\rm Ric}}\,.\label{epsilon4}
\ea

A comparison between the one-loop coefficients (which are not beta functions!) in the nonlocal theory \Eq{action} and the beta functions $\b_{2,3}$ of Stelle gravity gives \cite{Modesto:2022asj}
\be\label{betabeta}
\b_{\rm Ric} = \frac{\b_2}{(4\pi)^2} \, , \qquad \b_R = \frac{3 \b_3 - 2 \b_2}{6 (4 \pi)^2}\,. 
\ee
Therefore,
\be
\b_{\rm Ric}  +3 \b_R = \frac{3}{2}\frac{\b_3}{(4\pi)^2} \,,\label{b+3b}
\ee
and
\ba
\e_0 =  \frac{3 \e_2}{2}\frac{\b_3}{\b_2} \ll  \e_2 \,,\qquad \e_2-\e_0 = \e_2  \frac{2\b_2-3 \b_3}{2\b_2}\simeq \e_2 \,,\label{epsilon0}
\ea
where in \Eqq{epsilon0} we assumed that $\b_3 \ll \b_2$, as is usually the case since $\b_3$ depends on the number of fields $N_{\rm fields}$, while $\b_2$ does not (see \cite[formul\ae\ (9.1) and (9.8)]{BOS} and \cite[formula (4.137)]{Avramidi:2000bm}). From \Eqq{betabeta}, this implies
\be\label{bllb}
\frac{\b_{\rm Ric}  +3 \b_R}{\b_{\rm Ric}} \ll 1\,,
\ee
which can happen because $\b_R<0$.

In order to manipulate the propagator analytically, at this point we make the following approximations:
\ba
&& 1+ Q_2 = 1+ \ln \left(\frac{k^2}{\Lst^2} \right)^{\e_2} \simeq \left(\frac{k^2}{\Lst^2}\right)^{\e_2} \eqqcolon u^{2 \e_2}\,,\label{1pq1}\\
&& 1+ Q_0 = 1+ \ln \left(\frac{k^2}{\Lst^2} \right)^{\e_0}\simeq \left(\frac{k^2}{\Lst^2}\right)^{\e_0} = u^{2 \e_0}\,. \label{1pq2}
\ea
According to \Eqq{epsilon0}, $\e_0 \ll \e_2$, so that it is sufficient to focus on $\e_2$, the largest of these two parameters. For $\e_2 = - 0.0175$ (observed value), the approximation \Eq{1pq1} has an accuracy of 10\% or better if
\be
u^{2 \e_2} - (1+ Q_2) < 0.01 \quad \Longrightarrow \quad 0.02 < u < 60 \label{biggerInt} \,.
\ee
In the main text, we show that our cosmological model fully lies in this range of values. The other approximation we have to take into account is the self-consistency of the one-loop perturbative computation, namely,
\be
\left|\ln u^{2\e_2}\right| \ll 1\,,\qquad  \left|\ln u^{2\e_0}\right| \ll 1\, . \label{PertQG}
\ee
Again, for $\e_2 = - 0.0175$ and a quantum correction smaller than 10\% of the classical part, we found
\be
\left|\ln u^{2\e_2}\right| < 0.1 \quad \Longrightarrow \quad 0.06 < u < 17 \, .
\ee
This lower bound on $u$ is larger than the one in \Eq{biggerInt} and we can take it as a reference. For simplicity and to be even more conservative, we approximate it to
\be\label{z01}
u\gtrsim 0.1\,,
\ee
corresponding to an energy scale $E \gtrsim \Lst/10$, which is within the Weyl-invariant trans-Planckian regime if we identify $\Lst$ with $\Mpl$. 

Coming back to the calculation of the propagator, making use of the replacements \Eq{1pq1} and \Eq{1pq2} into \Eqq{PQEA3A2} we get
\ba
\bm{\cO}^{-1}(k) & \simeq & - \frac{4}{\rme^{\tilde{\g}_\textsc{e}} b}\frac{\Lz^2}{\Mpl^{2}k^4}
\left\{\frac{\bm{P}^{(2)}}{\left(\frac{k}{\Lst}\right)^{2\e_2}} + \frac{\bm{P}^{(0)}}{(D-2)^2 \left[\left(\frac{k}{\Lst}\right)^{2 \e_0}+ Q_b \right]}\right\}\nn
&=&-\frac{4}{\rme^{\tilde{\g}_\textsc{e}}b} \frac{\Lst^{2(1+\e_2)}}{\Mpl^{2}k^{4 + 2 \e_2 }}
\left[\bm{P}^{(2)} + \frac{\bm{P}^{(0)}}{(D-2)^2 \left(\frac{k}{\Lst}\right)^{2 (\e_0 - \e_2)}}\right].\label{PQEA3} 
\ea
Here, we assumed that $Q_b \ll 1$. Indeed, in $D=4$ dimensions
\be
Q_b = \frac{6\,\Lz^2 \, \b_R}{\rme^{\tilde{\g}_\textsc{e}} b \, \Mpl^2} \, \ln \de_0^2 =(\e_2-\e_0)\ln\de_0^2\,,
\ee
which is much smaller than 1 for $\e_2 - \e_0\simeq\e_2 \ll 1$ (see \Eqq{epsilon0}) or $\de_0\sim 1$. This analysis justifies the assumption $Q_b = 0$ in the main text.


\subsection{One-loop linearized equations of motion} \label{appeB4}

On the same vein, we can write the equations of motion \Eq{EoMQh} in vacuum as 
\ban
0&=& \B\left(\rme^{\bar{\H}} \left\{\left[P^{(2)} - (D-2) P^{(0)}\right]
+  \frac{\frac{2\b_{\rm Ric}}{\Mpl^2}\B\,\ln\left(-\frac{\B}{\Lst^2}\right)}{\exp\H_2}\, P^{(2)} \right. \right.\\
&& \left. \left.\hspace{1.5cm}
+ \frac{\B\left[8(D-1) \frac{\b_R}{\Mpl^2} \, \ln\left(-\frac{\B}{\Lst^2\,\de_0^2}\right)  + 2D \frac{\b_{\rm Ric}}{\Mpl^2} \, \ln\left(-\frac{\B}{\Lst^2}\right)\right]}{\exp\H_0}
P^{(0)}\right\}\right)_{\mu\nu}^{\rho\s}\,h_{\rho\s}\,,
\ean
where $\bar\H$ is taken at the zero order in the perturbation and, for $\Est\lesssim E \ll\Lst$, we can make the replacements \Eq{poly2} and \Eq{poly0}:
\ban
0 &=& \B \left(\rme^{\bar{\H}}\left\{\left[P^{(2)}-(D-2)P^{(0)}\right]
+\frac{\frac{2\b_{\rm Ric}}{\Mpl^2}\B \,\ln \left(-\frac{\B}{\Lst^2}\right)}{\rme^{\tilde{\g}_\textsc{e}}\left(-b\frac{\B}{\Lz^2}\right)}\, P^{(2)} \right.\right.\\
&& \left. \left.\hspace{1.5cm}
+ \frac{\B \left[8 (D-1) \frac{\b_R}{\Mpl^2} \,\ln \left(-\frac{\B}{\Lst^2\,\de_0^2}\right)  + 2D \frac{\b_{\rm Ric}}{\Mpl^2} \,\ln \left(-\frac{\B}{\Lst^2}\right)\right]}{\rme^{\tilde{\g}_\textsc{e}}b (D-2) \frac{\B}{\Lz^2}}P^{(0)}\right\}\right)_{\mu\nu}^{\rho \s} \, h_{\rho \s}\\
&=& \B \left(\rme^{\bar{\H}} \left\{\left[P^{(2)}-(D-2)P^{(0)}\right]
-  \frac{2\b_{\rm Ric} \Lz^2 \,\ln \left(-\frac{\B}{\Lst^2}\right)}{b \, \rme^{ \tilde{\g}_\textsc{e}} \Mpl^2}\, P^{(2)} \right. \right.\\
&& \left. \left.\hspace{1.5cm}
+ \Lz^2\frac{8(D-1)\b_R\ln \left(-\frac{\B}{\Lst^2 \,\de_0^2}\right)+2D \b_{\rm Ric} \, \ln \left(-\frac{\B}{\Lst^2}\right)}{ b \, \rme^{\tilde{\g}_\textsc{e}} (D-2) \Mpl^2}
P^{(0)}\right\}\right)_{\mu\nu}^{\rho \s} \, h_{\rho \s} \, .
\ean
Using again the completeness relations for the projectors \Eq{projectors}, we can factorize the operator $P^{(2)} -  (D - 2)  P^{(0)}$ and get
\ban
0&=& \B \, \left\{\rme^{ \bar{\H} } \left[P^{(2)} -(D-2)P^{(0)} \right]
\left[\mathbbm{1}-\frac{2\b_{\rm Ric}\Lz^2 \, \ln \left(-\frac{\B}{\Lst^2}\right)}{b \,\rme^{ \tilde{\g}_\textsc{e}}\Mpl^2} \, P^{(2)} \right. \right.\\
&& \left. \left.\hspace{1.5cm}
- \Lz^2 \frac{8 (D-1) \b_R \ln \left(-\frac{\B}{\Lst^2 \,\de_0^2}\right) 
	+ 2D \b_{\rm Ric} \,\ln \left(-\frac{\B}{\Lst^2}\right)}{ b \, \rme^{\tilde{\g}_\textsc{e}} (D-2)^2 \Mpl^2}P^{(0)}\right]\right\}_{\mu\nu}^{\rho \s} \, h_{\rho \s}\,,
\ean
where $\mathbbm{1}_{\mu\nu, \rho \s} \coloneqq (\eta_{\mu\rho} \eta_{\nu \s} + \eta_{\mu\s} \eta_{\nu \rho})/2$ is the identity tensor \cite{Accioly:2002tz}. Finally, for $Q_b \ll 1$ we write the linearized quantum equations of motion as 
\ba
0&=&\B \left\{\rme^{\bar{\H}} \left[P^{(2)}-(D-2)P^{(0)}\right]
\left[\mathbbm{1}+ \e_2 \, \ln \left(-\frac{\B}{\Lst^2}\right) \, P^{(2)} 
+ \e_0 \, \ln \left(-\frac{\B}{\Lst^2}\right) \,P^{(0)}\right]\right\}_{\mu\nu}^{\rho \s} \, h_{\rho \s}\nn
&=& \B \left\{\rme^{\bar{\H}} \left[P^{(2)}-(D-2)P^{(0)}\right]
\left[\mathbbm{1}+ Q_2 \,P^{(2)} + Q_0 \, P^{(0)}\right]\right\}_{\mu\nu}^{\rho \s} \, h_{\rho \s}\,.\label{EoMQh4}
\ea

In the presence of matter, the equations of motion are \Eq{LEOM}. Expanding them at the linear level around the Minkowski background (indeed, the metric $\eta_{\mu\nu}$ is an exact solutions of the equations of motion \Eq{LEOM} also including the quantum corrections $E^Q$), the contribution of the nonlocal form factor is of zero order in the fluctuations. Hence, it is diagonal and we can write the modified Einstein's equations \Eq{LEOM} as
\be
\left[\rme^{\H^{(0)}}\right]_{\mu\nu}^{\s\t}G^{(1)}_{\s\t} + \frac{2E^{Q {(1)}}_{\mu\nu}}{\Mpl^{2}}=\frac{1}{\Mpl^{2}}\left[\rme^{\H^{(0)}} \right]_{\mu\nu}^{\s\t} \, T^{(1)}_{\s\t}\,,
\label{LinEoM}
\ee
where $^{(0)}$ and $^{(1)}$ stand for, respectively, the zero-order and the first-order expansion in the metric or matter perturbations. There is no $\H^{(1)}$ contribution, as shown in \cite{Modesto:2021soh}. The left-hand side of \Eqq{LinEoM} at the linear level is given by \Eqq{EoMQh}. Hence, the equations of motion \Eq{LinEoM} at the linear level read
\ban
&&\B \left(\cancel{\rme^{\H^{(0)}}}\left\{\left[P^{(2)}-(D-2)P^{(0)}\right]+ \frac{\B \g_2^{Q}}{\rme^{\H_2^{(0)}}}\,P^{(2)}
+\frac{\B\left[4(D-1)\g_0^{Q}+D\g_2^{Q}\right]}{\rme^{\H_0^{(0)}}}P^{(0)}\right\}\right)_{\mu\nu}^{\s\t}h_{\s\t}\\
&&\qquad= \frac{1}{\Mpl^{2}}\big[\cancel{\rme^{\H^{(0)}}\big]_{\mu\nu}^{\s\t}}\,T^{(1)}_{\s\t}, 
\ean
and, factoring out the exponential form factor $\exp\H^{(0)}$ on both sides (which can be done since $\exp\bar\H$ is entire),
\ba
\frac{T^{(1)}_{\mu\nu}}{\Mpl^{2}}&=&\B\left\{\left[P^{(2)}-(D-2)P^{(0)}\right]+\rme^{-\H_2^{(0)}} \B \g_2^{Q}\, P^{(2)}\right.\nn
&& \qquad\left.
+\rme^{-\H_0^{(0)}} \B \left[4(D-1)\g_0^{Q}+D\g_2^{Q}\right]P^{(0)}\right\}_{\mu\nu}^{\s\t} h_{\s\t}\,, 
\label{EoMLinT}
\ea
where one should not confuse the labels $^{(0)}$ and $^{(2)}$ of the projectors with the linearization subscripts. Finally, using the result \Eq{EoMQh4} and a compact vectorial notation, we can replace the quantum form factors in \Eqq{EoMLinT} to end up with
\be
\frac{\bm{T}^{(1)}}{\Mpl^{2}}=\B\left[\bm{P}^{(2)}-(D-2)\bm{P}^{(0)}\right]\left[\bm{1}+ Q_2\bm{P}^{(2)} + Q_0 \bm{P}^{(0)}\right]\bm{h}\,.
\label{EoMQh10}
\ee
As stated in \cite{Modesto:2022asj}, 
 the linearized quantum equations of motion \Eq{EoMQh10} reduce to the linearized Einstein's equations in the classical limit  $\hbar \rightarrow 0$, i.e., for $Q_2=Q_0=0$.


\section{Spatial correlation function and power spectrum} \label{appeD}

The spatial correlation function for a generic field $\Psi(\bm{x})$ is defined by
\ba
\langle \Psi(\bm{x}) \Psi( \bm{y}) \rangle &=&
\langle \Psi (\bm{x}) \Psi(\bm{x} - \bm{r}) \rangle\nn
&\eqqcolon& \int \frac{\rmd^3 \bm{k}}{(2 \pi)^3}\, P_\Psi(k) \rme^{\rmi \bm{k} \cdot \bm{r}}\nn
&=& \int_0^{2\pi} \rmd\phi \int_0^{+\infty} \frac{\rmd k\, k^2}{(2 \pi)^3}\, P_\Psi(k) \int_{-1}^1 \rmd u  \, \rme^{\rmi k r u}\nn
&=& \int_0^{+\infty} \frac{\rmd k\, k^2}{(2 \pi)^2}\, P_\Psi(k) \int_{-1}^1 \rmd u \, \rme^{\rmi k r u}\nn
&=& \int_0^{+\infty} \frac{\rmd k\, k^2}{(2 \pi)^2}\, P_\Psi(k) \frac{2\sin(kr)}{kr}\nn
&=& \int_0^{+\infty} \frac{\rmd k\, k^2}{2 \pi^2}\, P_\Psi(k) \frac{\sin(kr)}{kr} \, , 
\label{PGk}
\ea
where $\bm{r} = \bm{x} - \bm{y}$, $r = |\bm{r}|$, and $k = |\bm{k}|$. It is common to define the dimensionless power spectrum as
\be
\Delta^2_\Psi(k) \coloneqq \frac{k^3}{2 \pi^2}\,P_\Psi(k)\, , 
\label{DeltaG}
\ee
so that \Eqq{PGk} turns into
\be\label{psipsiDe}
\langle \Psi(\bm{x}) \Psi( \bm{y}) \rangle =  \int_0^{+\infty}\frac{\rmd k}{k}\,\Delta^2_\Psi(k)\,\frac{\sin(kr)}{kr}\,.
\ee


\section{Two-point correlation function of the energy-momentum tensor}\label{appeE}

In this section, we provide all the details about the computation of the two-point correlation function of the perturbed energy-momentum tensor $T_{\mu\nu}^{(1)}$. For the sake of simplicity, we omit the indices, which are hidden in the boldface notation. Hence, from the linearized equations of motion \Eq{EoMQh10} we can get the two-point function
\ba
\frac{1}{\Mpl^{4}} \langle \bm{T}^{(1)} (x) \,  \bm{T}^{(1)} (y) \rangle \!\!&=&\!\!\B_x\left[\bm{P}_x^{(2)}-(D-2)\bm{P}_x^{(0)}\right]\left[\bm{1}+ Q_2\bm{P}_x^{(2)} + Q_0 \bm{P}_x^{(0)} \right] \nn
&&\times
\B_y   \left[\bm{P}_y^{(2)} -  (D - 2)  \bm{P}_y^{(0)} \right]\left[\bm{1}+ Q_2\bm{P}^{(2)}_y + Q_0 \bm{P}^{(0)}_y \right]\langle \bm{h}(x) \bm{h}(y) \rangle \nn
&=&\!\! \int \!\frac{\rmd^D k}{(2\pi)^D}\,\rme^{\rmi k\cdot(x -y)}\bm{\cO}^{-1}(k)\,k^4\nn
&&\times\left[\bm{P}^{(2)}-(D-2)\bm{P}^{(0)}\right]\left[\bm{1}+ Q_2\bm{P}^{(2)} + Q_0 \bm{P}^{(0)} \right] \nn
&& \times
\left[\bm{P}^{(2)} -  (D - 2)  \bm{P}^{(0)} \right]\left[\bm{1}+ Q_2\bm{P}^{(2)} + Q_0 \bm{P}^{(0)} \right]\nn
&\stackrel{\textrm{\tiny \Eq{PQEA3A2}}}{=}& \!\!\int \!\frac{\rmd^D k}{(2\pi)^D}\,\rme^{\rmi k\cdot(x -y)}\, k^4 \left[\bm{P}^{(2)} -  (D - 2)  \bm{P}^{(0)} \right]\left[\bm{1}+ Q_2\bm{P}^{(2)} + Q_0 \bm{P}^{(0)} \right]\nn
&&\times \left[\bm{P}^{(2)} -  (D - 2)  \bm{P}^{(0)} \right]\left[\bm{1}+ Q_2\bm{P}^{(2)} + Q_0 \bm{P}^{(0)} \right]\nn
&& \times 
\left(-\frac{4}{b\,\rme^{\tilde{\g}_\textsc{e}}}\frac{\Lz^2}{\Mpl^{2}k^4}\right) \!\left[\frac{\bm{P}^{(2)}}{1+ Q_2} + \frac{\bm{P}^{(0)}}{ (D- 2)^2  \left(1+ Q_0  + Q_b \right)}\right].\label{scalarP}
\ea
Making iterated use of the properties \Eq{copro} in the product of the first and third brackets with the last bracket, we can further simplify \Eqq{scalarP},  
\ba
\langle \bm{T}^{(1)} (x) \,  \bm{T}^{(1)} (y) \rangle &=&-\frac{4\Mpl^{2} \Lz^2}{b \, \rme^{\tilde\g_\textsc{e}}} \int \! \frac{\rmd^D k}{(2\pi)^D}\,\rme^{\rmi k\cdot (x -y)} \left[\bm{1} + Q_2 \bm{P}^{(2)} + Q_0 \bm{P}^{(0)} \right]\nn
&&\times\left[\bm{1} + Q_2 \bm{P}^{(2)} + Q_0 \bm{P}^{(0)} \right]\left[\frac{\bm{P}^{(2)}}{1+ Q_2}+ \frac{\bm{P}^{(0)}}{1+ Q_0}\right]\nn
&=& -\frac{4\Mpl^{2} \Lz^2}{b \, \rme^{\tilde\g_\textsc{e}}} \int \! \frac{\rmd^D k}{(2\pi)^D}\,\rme^{\rmi k\cdot (x -y)} \left[\bm{1}+ (2 Q_2  + Q_2^2)  \bm{P}^{(2)} + (2 Q_0 + Q_0^2) \bm{P}^{(0)} \right]\nn
&&\times \left[\frac{\bm{P}^{(2)}}{1+ Q_2}+\frac{\bm{P}^{(0)}}{1+Q_0}\right]\nn
&=&-\frac{4\Mpl^{2} \Lz^2}{b \, \rme^{\tilde\g_\textsc{e}}} \int \! \frac{\rmd^D k}{(2\pi)^D}\,\rme^{\rmi k\cdot (x-y)}\left[\frac{(1+Q_2)^2\bm{P}^{(2)}}{1+ Q_2}+ \frac{(1+Q_0)^2\bm{P}^{(0)}}{1+Q_0}\right]\nn
&=&-\frac{4\Mpl^{2} \Lz^2}{b \, \rme^{\tilde\g_\textsc{e}}} \int \! \frac{\rmd^D k}{(2\pi)^D}\,\rme^{\rmi k\cdot (x-y)}\left[(1+ Q_2)\bm{P}^{(2)}+(1+Q_0)\bm{P}^{(0)}\right]\,. 
\label{TT}
\ea
Let us now display the indices and also introduce the explicit form of the projectors \Eq{projectors} in the two-point function \Eq{TT}: 
\ba
\langle  T^{(1)}_{\mu\nu} (x) \,  T^{(1)}_{\rho \s} (y) \rangle  
& = &-\frac{4\Mpl^{2} \Lz^2}{b \, \rme^{\tilde\g_\textsc{e}}}  
\int \! \frac{\rmd^D k}{(2\pi)^D}\,\rme^{\rmi k\cdot(x-y)}\left[(1+ Q_2) P_{\mu\nu\rho\s}^{(2)} + (1+Q_0) P_{\mu\nu\rho\s}^{(0)}\right] \nn
& \stackrel{\textrm{\tiny \Eq{projectors}}}{=}& 
-\frac{4\Mpl^{2} \Lz^2}{b \, \rme^{\tilde\g_\textsc{e}}} \int \! \frac{\rmd^D k}{(2\pi)^D}\,\rme^{\rmi k\cdot(x-y)}\nn
&&\times\left\{(1+ Q_2) \left[  \frac{1}{2} \left(\Theta_{\mu \rho} \Theta_{\nu \s} +  \Theta_{\mu\s} \Theta_{\nu\rho} \right)- \frac{1}{D-1} \Theta_{\mu\nu} \Theta_{\rho\s} \right]\right.\nn
&&
\qquad
\left.  +(1+Q_0) \frac{1}{D-1} \Theta_{\mu \nu} \, \Theta_{\rho \s} \right\}\nn
&=& -\frac{4\Mpl^{2} \Lz^2}{b \, \rme^{\tilde\g_\textsc{e}}} 
\int \! \frac{\rmd^D k}{(2\pi)^D}\,\rme^{\rmi k\cdot(x-y)}\nn
&&\times
\left((1+ Q_2) \left\{\frac{1}{2} \left[\left(\eta_{\mu \rho} - \frac{k_\mu k_\rho}{k^2} \right)\left(\eta_{\nu \s} - \frac{k_\nu k_\s}{k^2} \right)\right. \right. \right. \nn
&&\qquad\qquad \left. 
+  \left(\eta_{\mu \s} - \frac{k_\mu k_\s}{k^2} \right)
\left(\eta_{\nu \rho} - \frac{k_\nu k_\rho}{k^2} \right)
\right] \nn
&&\qquad\qquad\left.- \frac{1}{D-1} \left(\eta_{\mu \nu} - \frac{k_\mu k_\nu}{k^2} \right) 
\left(\eta_{\rho \s} - \frac{k_\rho k_\s}{k^2} \right)\right\}\nn
&&\qquad\qquad\left. + (1+Q_0) \frac{1}{D-1}  \left(\eta_{\mu \nu} - \frac{k_\mu k_\nu}{k^2} \right)
\left(\eta_{\rho \s} - \frac{k_\rho k_\s}{k^2} \right) \right) \, . 
\ea
Expanding the products in the brackets:
\ba
\langle  T^{(1)}_{\mu\nu} (x) \,  T^{(1)}_{\rho \s} (y) \rangle  
&=&  -\frac{4\Mpl^{2} \Lz^2}{b \, \rme^{\tilde\g_\textsc{e}}} \int \! \frac{\rmd^D k}{(2\pi )^D}\,\rme^{\rmi k\cdot (x-y)}\nn
&&\qquad\times\left\{(1+ Q_2) \left[\frac{1}{2} \left(\eta_{\mu \rho} \eta_{\nu \s} - \eta_{\mu \rho} \frac{k_\nu k_\s}{k^2}  - \frac{k_\mu k_\rho}{k^2} \eta_{\nu \s}
+  \frac{k_\mu k_\rho k_\nu k_\s}{k^4}\right. \right. \right. \nn
&& \quad\qquad\left. \left. \left. 
+  \eta_{\mu \s} \eta_{\nu \rho}- \eta_{\mu \s} \frac{k_\nu k_\rho}{k^2} 
- \frac{k_\mu k_\s}{k^2} \eta_{\nu \rho}+ \frac{k_\mu k_\s k_\nu k_\rho}{k^4}  \right) 
\right. \right.\nn
&& \quad\qquad \left. - \frac{1}{D-1} \left(\eta_{\mu \nu} \eta_{\rho \s} 
- \eta_{\mu \nu} \frac{k_\rho k_\s}{k^2}- \frac{k_\mu k_\nu}{k^2}  \eta_{\rho \s} 
+ \frac{k_\mu k_\nu k_\rho k_\s}{k^4} \right) \right]\nn
&&\quad\qquad\left. +(1+Q_0)  \frac{1}{D-1}  \left(\eta_{\mu \nu} \eta_{\rho \s} 
- \eta_{\mu \nu} \frac{k_\rho k_\s}{k^2}- \frac{k_\mu k_\nu}{k^2}  \eta_{\rho \s} 
+ \frac{k_\mu k_\nu k_\rho k_\s}{k^4} \right)\right\}.\nn \label{TT2}
\ea
According to Lorentz invariance, the following identities hold:
\ba
&& \int \rmd^D k \, f(k) \, k_\mu k_\nu = \frac{1}{D} \, \eta_{\mu\nu} \int \rmd^D k \, f(k) \, k^2 \, , \\
&& \int \rmd^D k \, f(k) \, k_\mu k_\nu k_\rho k_\s = \frac{1}{D(D+2)}  
\left( \eta_{\mu\nu}  \eta_{\rho \s} 
+ \eta_{\mu\rho}  \eta_{\nu \s}
+  \eta_{\mu\s}  \eta_{\nu \rho }
\right) \int \rmd^D k \, f(k) \, k^4 \,,
\ea
which can be easily verified contracting the left-hand sides with the Minkowski metric. Applying these identities to \Eqq{TT2}, we obtain
\ba
\langle  T^{(1)}_{\mu\nu} (x) \,  T^{(1)}_{\rho \s} (y) \rangle  
& = &-\frac{4\Mpl^{2} \Lz^2}{b \, \rme^{\tilde\g_\textsc{e}}}  
\int \! \frac{\rmd^D k}{(2\pi)^D}\,\rme^{\rmi k\cdot(x-y)}\nn
&&\times\left\{(1+ Q_2) \frac{(D-2)(D+1)}{(D+2)(D-1)}\left[\frac12\left(\eta_{\mu\rho}\eta_{\nu\s}+\eta_{\mu\s}\eta_{\nu\rho}\right)-\frac{1}{D}\eta_{\mu\nu}\eta_{\rho\s}\right]\right.\nn
&&\qquad\left. + (1+Q_0)\frac{1}{D(D+2)(D-1)}\left[(D^2-3)\eta_{\mu\nu}\eta_{\rho\s}+\eta_{\mu\rho}\eta_{\nu\s}+\eta_{\mu\s}\eta_{\nu\rho}\right]\right\}.\nn
\ea
Therefore, the two-point correlation function of the perturbation of the energy density in $D=4$ dimensions is
\ba
\langle \de \rho(x)\de\rho(y)\rangle &=& \langle T^{(1)}_{00} ({x}) T^{(1)}_{00} ({y}) \rangle\nn
&=& -\frac{4\Mpl^2 \Lz^2}{b \, \rme^{\tilde\g_\textsc{e}}}  
\int \! \frac{\rmd^D k}{(2\pi)^D}\,\rme^{\rmi k\cdot(x-y)}\frac{D+1}{D(D+2)}\left[(D-2)(1+ Q_2) + (1+Q_0)\right]\nn
&\stackrel{D=4}{=}&-\frac{4\Mpl^2 \Lz^2}{b \, \rme^{\tilde\g_\textsc{e}}}  
\int \! \frac{\rmd^4 k}{(2\pi)^4}\,\rme^{\rmi k\cdot(x-y)}\frac{5}{24}\left[2(1+ Q_2)+ (1+Q_0)\right].\label{roro}
\ea
Replacing \Eqqs{1pq1} and \Eq{1pq2} into \Eqq{roro}, we find
\ba
\langle\de\rho(x)\de\rho(y) \rangle & = & -\frac{4\Mpl^2\Lz^2}{b\,\rme^{\tilde\g_\textsc{e}}} \int \! \frac{\rmd^4 k}{(2\pi)^4}\,\rme^{\rmi k\cdot(x-y)}\frac{5}{24}\left[2\left(\frac{k}{\Lst}\right)^{2\e_2}+\left(\frac{k}{\Lst}\right)^{2\e_0}\right].\label{roro2}
\ea



\begin{thebibliography}{99}
\bibitem{Heisenberg:2022gqk} \au{L}{Heisenberg}, \au{H}{Villarrubia-Rojo} and \au{J}{Zosso}, \tiaq{Can late-time extensions solve the $H_0$ and $\sigma_8$ tensions?} \arX{2202.01202}.
\end{thebibliography}
\end{document}